\documentclass[final,5p,times,twocolumn]{elsarticle}

\usepackage{amssymb}
\usepackage{tensor}

\usepackage{amsmath}
\usepackage{xcolor}
\usepackage{subcaption}
\usepackage{dsfont}
\usepackage{mhchem}
\usepackage{background}

\journal{arXiv}

\usepackage{hyperref}
\hypersetup{
     colorlinks=true,
      linkcolor=blue,
}

\backgroundsetup{contents=}
\begin{document}

\begin{frontmatter}



\title{A laser-driven mixed fuel nuclear fusion
micro-reactor concept}


\author{Hartmut Ruhl and Georg Korn}

\address{Marvel Fusion, Blumenstrasse 28, Munich, Germany}

\begin{abstract}
We propose a laser-driven near-solid density nano-structured
micro-reactor concept operating with mixed nuclear fusion fuels.
The micro-reactor is capable of making use of a range 
of neutronic and aneutronic fuels. Its core parts consists of an 
embedded nanoscopic nuclear fuel based laser-driven 
nano-accelerator that is capable of producing non-thermal fuel
distributions almost instantly.
\end{abstract}

\begin{keyword}
integrated accelerator, nanoscopic reactor, nanoscopic converter,
nonlinear optics, secular field generator, non-thermal Lawson criteria



\end{keyword}

\end{frontmatter}


\tableofcontents

\section{Introduction to the concept}
\label{sect_introduction}
In recent years there has been an abundance of papers in the field of
ultra-short ultra-intense laser-matter interaction with
nano-structures. We quote \cite{fedeli2018ultra,
margarone2022target,curtis2018micro} and the literature
therein to give examples.

Since lasers represent the fastest macroscopic energy sources we
propose a micro-reactor powered by ultra-short ultra-high intensity
laser pulses with UV to VUV wavelengths. Nano-structures
promise an efficient way of transferring laser energy into a target.
Simulations show that near complete laser deposition in 
nano-structured targets is possible.

With the help of the conversion fraction $\eta^{kl}$, which is defined
later in the paper, modes of operation of the micro-reactor and their
limits can be discussed and linked to existing experiments. In
addition, with the help of $\eta^{kl}$ it is possible to identify
fusion conversion enhancing configurations.

Many of the papers in the field discuss nano-structures that are
either extremely small and randomly oriented or the structures are so
large that they cannot be called nano-structures anymore.
Here, we propose a nano-structured nuclear micro-reactor with
comparatively small structures size, which is entirely composed of
nuclear fuels.

It comprises an integrated nano-structured accelerator
consisting of boron or lithium rods with structure radii of $R \le 30
\, \text{nm}$ doped with constituents of nuclear fuels.
The nuclear fuels can have a range of Gamov energies and
$S$-factors. They can be neutronic or aneutronic or any mix of the
latter. There are interesting chemical compounds like lithium
borohydrate, that have very high natural densities for these nuclear
fuel constituents.

The nano-rods are supposed to ionize rapidly leading to
subsequent fuel releasing Coulomb explosions propagating close to the
speed of light along the laser pulse. Coulomb explosions are efficient
if electronic recurrence into the ionizing nano-structures is
avoided. In addition, ultra-short high energy laser pulses are capable of
over-heating electrons thus reducing electronic collisionality.

The purpose of the present paper is to outline the nano-structured
micro-reactor concept, the abstraction model, and to discuss some of its
properties, modes of operation, and its limitations on a parametric
level.

The paper is structured in the following way. In section
\ref{concept} the abstraction model is outlined for future
reference. In section \ref{lawson} simplified relativistic transport
equations are discussed in preparation of the numerical model
in future versions of this paper. With the help of the relativistic
transport equations the parameters required for high fusion efficiency
can be identified. In section \ref{nano_accelerator} the concept of
the embedded nano-accelerator is introduced. In
section \ref{burn_fraction} the conversion fraction and
efficiency are addressed. In section \ref{nonlinear_optics} the laser
parameters required for the desired optical properties of the laser
pulses interacting with nano-rods are sketched. In section
\ref{collisions_radiation} radiative energy loss processes are
addressed.

\section{The abstraction model} \label{concept}
Since the required laser pulse radiation is ultra-intense electrons
will become relativistic. There might also be positrons. Hence, we refer to
\cite{van1973elements,VASAK1987462,ZHUANG1996311,van1978derivation}.
For furture reference we state an expansion into a quantum
BBGKY-hierarchy up to binary correlation order in the presence of
electromagnetic fields leads to
\begin{eqnarray}
\label{eq:trans-on-shell-new}
&&\left( p^{\mu}_k \frac{\partial}{\partial x^{\mu}} + m_k F^{i}_k 
   \frac{\partial}{\partial {p^{i}_k}} \right) \, f \left( x, \vec{p}_k \right) \\
&&=\sum_{l, k^{\, '} l^{\,'}} \int \frac{d^3p_l}{p^0_l} \frac{d^3p^{\,'}_{l^{\,
   '}}}{p^{\,' 0}_{l^{\, '}}}
   \frac{d^3p^{\,'}_{k^{\, '}}}{p^{\,' 0}_{k^{\, '}}} \, \nonumber \\
&&\hspace{1.5cm} \times \, {\cal A} \left( p_l \, p_k, p^{\,'}_{k^{\, '}}
   p^{\,'}_{l^{\, '}} \right)  \,
   f \left( x, \vec{p}^{\,'}_{l^{\, '}} \right) \,
   f \left( x, \vec{p}^{\,'}_{k^{\, '}} \right) \nonumber \\
&&\hspace{0.5cm} - \sum_{l, k^{\, '} l^{\,'}} \int \frac{d^3p_l}{p^0_l} \,
   \frac{d^3p^{\,'}_{k^{\, '}}}{p^{\,' 0}_{k^{\, '}}}
   \frac{d^3p^{\,'}_{l^{\, '}}}{p^{\,' 0}_{l^{\, '}}} \, \nonumber \\
&&\hspace{1.5cm} \times \, {\cal A} \left( p^{\,'}_{l^{\, '}} p^{\,'}_{k^{\, '}}, p_k \, p_l \right)
   \, f \left( x, \vec{p}_k \right) \,
   f \left( x, \vec{p}_l \right) \, , \nonumber
\end{eqnarray}
where the invariant transition amplitude is given by
\begin{eqnarray}
\label{eq:trans-rate}
  &&{\cal A} \left( p_l \, p_k, p^{\,'}_{k^{\, '}}
   p^{\,'}_{l^{\, '}} \right) \\
  &&= \delta^4 \left( p_l+p_k- p^{\,'}_{l^{\, '}}- p^{\,'}_{k^{\, '}} \right) \, \left| \left\langle p_l \, p_k \, \middle| {\cal T}_{in}
            \middle| p^{\,'}_{k^{\, '}} p^{\,'}_{l^{\, '}} \right\rangle \right|^2 \, . \nonumber
\end{eqnarray}
The binary ${\cal T}_{in}$-matrix in (\ref{eq:trans-rate}) has to be
calculated in the context of ultra-strong electromagnetic
fields. Hence, appropriately dressed states are required. Calculations
of that kind in a somewhat different context are found in
\cite{king2013trident}. The ${\cal T}_{in}$-matrix is obtained with
the help of the $\tensor{{\bf S}}{_{in}}$-matrix, which is to lowest order
\begin{eqnarray}
\label{eq:Smatrix-low}
   &&\tensor{{\bf S}}{_{in}}=\mathds{1}+\frac{1}{\hbar c} \, \int^{\infty}_{-\infty}
      d^4x \, :{\cal L}^{in}_I \left( x \right): \, .
\end{eqnarray}
This implies
\begin{eqnarray}
\label{eq:transition}
  &&\tensor[_{in}]{\left\langle q_1 q_2 \middle| {\bf S} - \mathds{1} \middle| p_2 p_1
      \right\rangle}{_{in}} \\
  &&= \frac{i}{\hbar c} \, \left( 2 \pi \hbar
     \right)^4 \, \delta^4 \left( p_1 + p_2 - q_1 -q_2 \right) \,
     \nonumber \\
  &&\hspace{1cm} \times \,  \tensor[_{in}]{\left\langle q_1 q_2 \middle| :{\cal L}^{in}_I: \middle| p_2 p_1
      \right\rangle}{_{in}} \nonumber
\end{eqnarray}
and hence
\begin{eqnarray}
\label{eq:t-matrix}
  &&\left\langle q_1 q_2 \middle| {\cal T} \middle| p_2 p_1
      \right\rangle \\
  &&= \frac{2\pi}{c} \, \left( 2 \pi \hbar
      \right)^4 \, \tensor[_{in}]{\left\langle q_1 q_2 \middle| :{\cal
      L}^{in}_I: \middle| p_2 p_1 \right\rangle}{_{in}} \, . \nonumber
\end{eqnarray}

In section \ref{lawson} we give details of the abstraction model based
on the general relativist structure of the transport equations stated here.

\section{Rate equations for reactions} \label{lawson}
\begin{figure}
  \begin{center}
  \includegraphics[width=80mm]{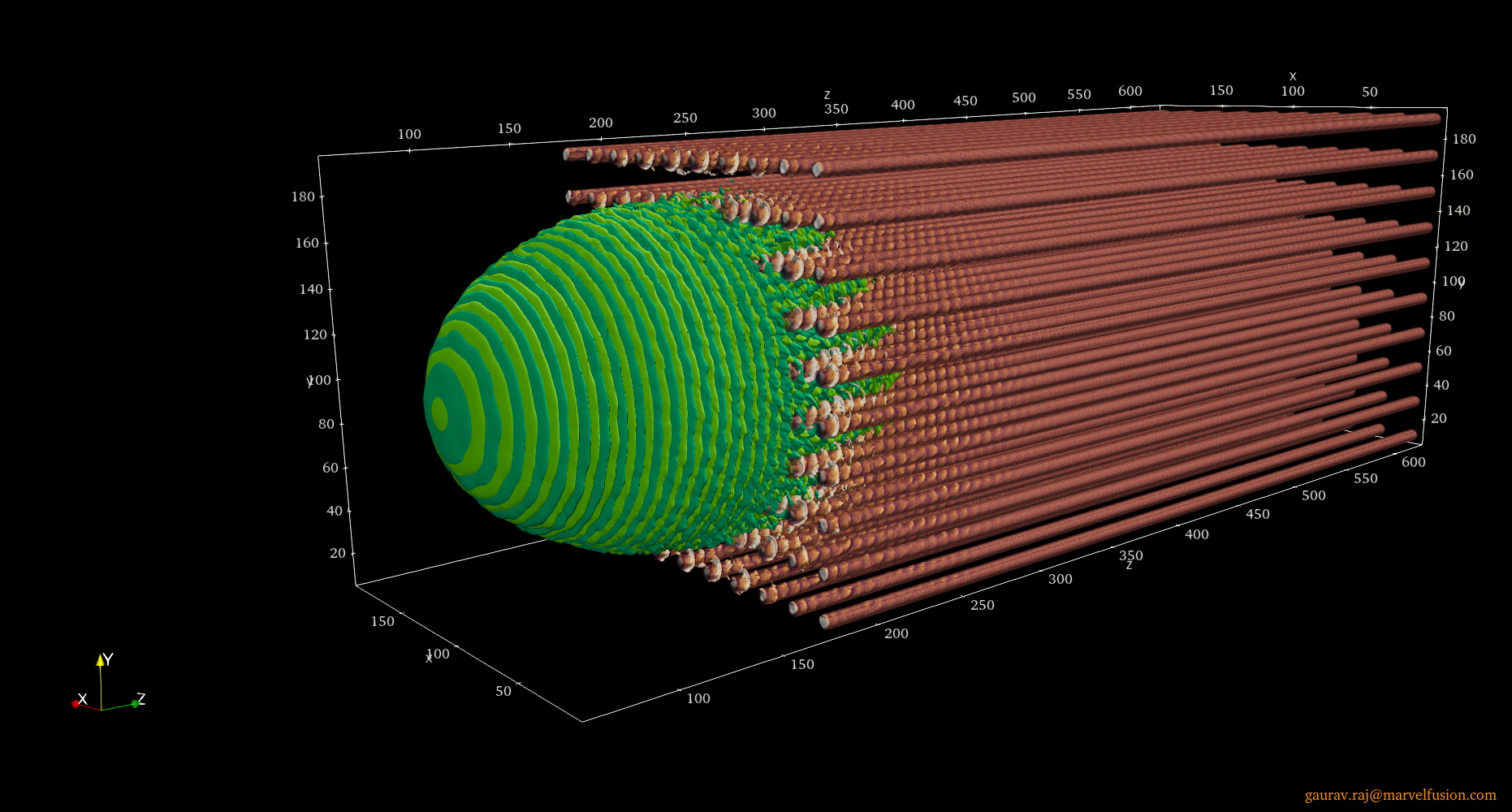}
  \caption{\label{absorber} Cylindrical boron nano-rods with $R \le 30 
    \, \text{nm}$ are placed inside the micro-reactor. The boron 
    nano-rods contain protons, deuterons, and tritium. An ultra-short
    high intensity UV laser pulse impinges on the reactor from the left
    and couples to the boron nano-rods. The nano-structured absorber
    extends beyond the laser focal spot in lateral directions.}
  \end{center}
\end{figure}

The abstraction model (\ref{eq:trans-on-shell-new}) -
(\ref{eq:t-matrix}) contains the dynamics of all particles
embedded in a self-consistent electromagnetic field context inside the
convertor as depicted in Fig. \ref{absorber}. It is also the basis of
the numerical abstraction model in future papers. This is one reason
why we state the details of it here. We will also make use of it later
in the paper.

A fusing $kl$-system can be approximately described by the following
kinetic equations
\begin{eqnarray}
\label{kl-system}
&&p^{\mu}_k \frac{\partial f_k}{\partial {x^{\mu}}} + m_k F^{i}_k 
   \frac{\partial f_k}{\partial {p^{i}_k}}\\
&=&\sum_{l, k^{\,'}l^{\,'}} \int \frac{d^3p_l}{p^0_l}   \int \frac{d^3p^{\,'}_{k^{\,
    '}}}{p^{\, ' 0}_{k^{\, '}}}  \int \frac{d^3p^{\,'}_{l^{\, '}}}{p^{\,' 0}_{l^{\, '}}}  \nonumber \\
&&\hspace{1cm} \times\left( W^C_{kl \, k^{'}l^{'}}\, f_{k^{\, '}}f_{l^{\, '}} -
   W^C_{k^{'}l^{'} kl}\, f_kf_l \right) \nonumber \\
&&-\sum_{{l, k^{'}, l^{ '}}} \int \frac{d^3p_l}{p^0_l}   \int \frac{d^3p^{\,'}_{k^{\,
    '}}}{p^{\,' 0}_{k^{\, '}}}  \int \frac{d^3p^{\,'}_{l^{\, '}}}{p^{\,' 0}_{l^{\, '}}} \,
    W^R_{k^{'}l^{'} kl}\, f_kf_l \nonumber 
\end{eqnarray}
and
\begin{eqnarray}
\label{force-eqns}
&&F^{\mu}_k = \frac{q_k}{m_k c} F^{\mu \nu} p_{k\nu} \, ,
\end{eqnarray}
where the $q_k$ are the electric charges of particles $k$ and $F^{\mu \nu}$ 
is the electric field strength tensor. Maxwell's equations are given by
\begin{eqnarray}
\label{maxwell-eqns}
&&\frac{\partial}{\partial {x^{\mu}}} F^{\mu \nu} =
\frac{j^{\nu}}{c^2 \epsilon_0} \; , \quad \frac{\partial}{\partial {x^{\mu}}} 
\tilde{F}^{\mu \nu} = 0 \, ,
\end{eqnarray}
where $\tilde{F}^{\mu \nu}$ is the dual of $F^{\mu \nu}$. The total
four current is
\begin{eqnarray}
\label{current-densities}
&&j^{\nu} = \sum_{k} q_k \int d^4p \; 2 \Theta (p_0) \; 
\delta (p^2 -m^2_kc^2) \; cp^{\nu} f_k \, ,
\end{eqnarray}
where the collisional and reactive invariant transition amplitudes
$W^C$ and $W^R$ can be mapped onto invariant collisional and reactive
cross sections $\sigma_C$ and $\sigma_R$ provided the underlying
system is sufficiently dilute and weakly coupling with the help of
\begin{eqnarray}
\label{trans-ampl}
&&\hspace{-1cm}W^C_{kl \, k^{'}l^{'}} = W^C_{k^{'}l^{'}, kl} \, , \\
&&\hspace{-1cm}W^C_{k^{'}l^{'} kl} =s \, \sigma^{kl}_C(s,\psi) \, 
\delta^4 \left( p^{\,'}_{k^{\,'}}+p^{\,'}_{l^{\,'}}-p_k -p_l \right)
\, \delta_{k^{\, '}k} \, \delta_{l^{\,'} l} \, , \\
&&\hspace{-1cm}W^R_{k^{\;'}l^{\;'}kl} = s \, \sigma^{kl}_R(s,\psi) \, 
\delta^4 \left( p^{\,'}_{k^{\,'}}+p^{\,'}_{l^{\,'}}-p_k -p_l \right) \, , \\
&&\hspace{-1cm}s=\left( p^{\mu}_k + p^{\mu}_l \right)^2 \, .
\nonumber
\end{eqnarray}
The kinematics of the reactions in the $kl$-system is best analyzed in
the center of mass frame
\begin{eqnarray}
\label{beta}
\vec{p}^{\; cm}_k&=&\vec{p}_k +\frac{1}{\beta^2} \, (\gamma -1) \, (\vec{p}_k \cdot 
\vec{\beta}) \, \vec{\beta} -\gamma \, \vec{\beta} \, p^0_{k} \, , \\
\vec{p}^{\; cm}_l&=&\vec{p}_l +\frac{1}{\beta^2} \, (\gamma -1) \, (\vec{p}_l \cdot 
\vec{\beta}) \, \vec{\beta} -\gamma \, \vec{\beta} \, p^0_{l} \, , \nonumber
\end{eqnarray}
where
\begin{eqnarray}
&&\vec{\beta} = \frac{\vec{p}_k+\vec{p}_l}{p^0_{k}+p^0_{l}} \, , \\
&&\gamma = \frac{1}{\sqrt{1-\beta^2}} \, , \\
&&p^0_{k}=\sqrt{m^2_kc^2 + \vec{p}^{\, 2}_k} \, , \\
&&p^0_{l}=\sqrt{m^2_lc^2 + \vec{p}^{\, 2}_l} \, .
\end{eqnarray}
The quantity $p$ is the length of the CM-frame momenta
$\vec{p}^{\, cm}_k$ and $\vec{p}^{\, cm}_l$. It is given by
\begin{eqnarray} 
p&=&\frac{1}{2\sqrt{s}} \, \sqrt{\left( s-(m^2_k+m^2_l) \, c^2 
\right)^2-4 \, m^2_km^2_l \, c^4} \; .  
\end{eqnarray} 
To define the post-collision momenta we introduce a right-handed  
coordinate system in the CM frame, the $\vec{e}_1$-axis of which is 
along $\vec{p}^{\; cm}_k$. The CM frame coordinate system is embedded
into a right-handed coordinate system in the lab frame. The $z$-axis of
the latter is along $\vec{e}_z$.
\begin{figure}[ht]
\begin{center}
\includegraphics[width=25mm]{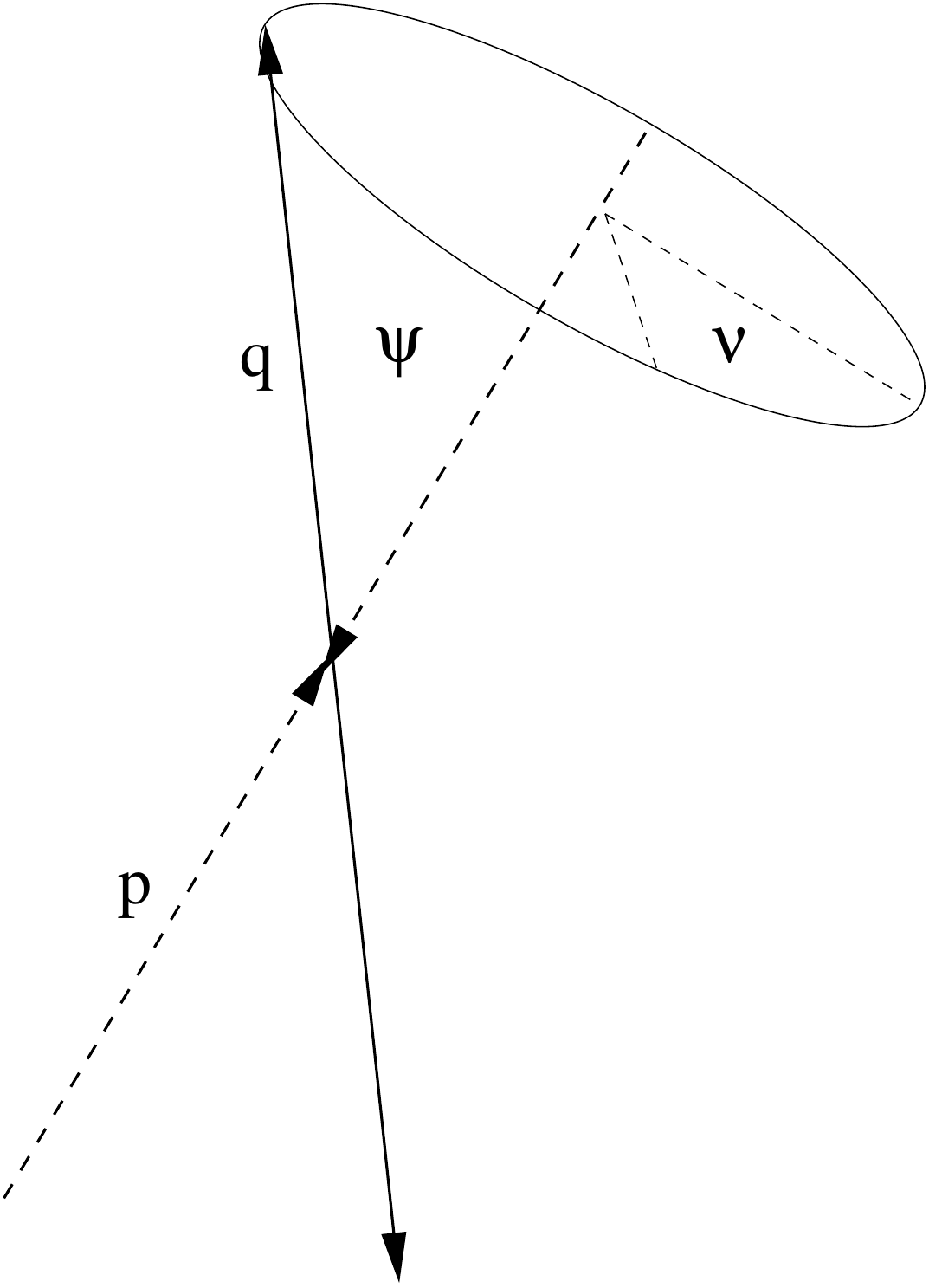}
\end{center}
\caption{\label{scattering} Release angles of nuclear fusion products
for binary decays.}
\end{figure}
For the parametrization of the nuclear fusion products in the
$k^{\, '}l^{\, '}$-system the angles $\psi$ and $\nu$ are introduced as
indicated in Fig. \ref{scattering}. We construct three orthonormal
vectors for all three coordinate axes as follows
\begin{eqnarray}
\label{right-handed-coord}
&&\vec{e}_1=\frac{\vec{p}^{\; cm}_k}{|\vec{p}^{\; cm}_k|} \, , \\
&&\vec{e}_2=\frac{\vec{p}^{\; cm}_k \times \vec{e}_z}{|\vec{p}^{\;  cm}_k 
   \times \vec{e}_z|} \, , \\
&&\vec{e}_3=\frac{(\vec{p}^{\; cm}_k \times 
\vec{e}_z) \times \vec{p}^{\;  cm}_k}{|(\vec{p}^{\; cm}_k \times \vec{e}_z) 
\times \vec{p}^{\; cm}_k|} \, .
\end{eqnarray}
If $s \ge (m_{k^{\, '}} c +m_{l^{\, '}}c)^2$ holds, 
where $m_{k^{\, '}}$ and $m_{l^{\, '}}$  denote either the
post-collisional masses or the masses of the binary nuclear fusion
products, we can calculate the post-collisional or post-fusion
momenta in the CM-frame
\begin{eqnarray}
\label{CM-momentum-vec-q}
&&\hspace{-1cm}p^{\, ' 0 \; cm}_{k^{\, '}}=\sqrt{m^2_{k^{\, '}} c^2 + q^2} \; , \\
&&\hspace{-1cm}\vec{p}^{\, ' \; cm}_{k^{\, '}}=q \, \cos \psi \vec{e}_1 + q \, \sin \psi
\, \sin \nu \, \vec{e}_2 + q \, \sin \psi \, \cos \nu \, 
\vec{e}_3 \; , \\
&&\hspace{-1cm}p^{\, ' 0 \; cm}_{l^{\, '}}= \sqrt{m^2_{l^{\, '}} c^2 + q^2} \; , \\
&&\hspace{-1cm}\vec{p}^{\, ' \; cm}_{l^{\, '}} =-\vec{p}^{\, ' \; cm}_{k^{\, '}} \; ,
\end{eqnarray}
where $q$ is given by 
\begin{eqnarray}
\label{CM-momentum-q}
q&=&\frac{1}{2\sqrt{s}} \, \sqrt{\left( s-(m^2_{k^{\, '}}+m^2_{l^{\, '}}) \, c^2 
\right)^2-4 \, m^2_{k^{\, '}}m^2_{l^{\, '}} \, c^4} \; . 
\end{eqnarray}
Finally we transform back into the lab-frame. Since the CM frame 
moves with the velocity $c\vec{\beta}$ we can go back to the lab frame 
by boosting the CM frame with the velocity $-c \vec{\beta}$. 
We obtain for the post-collision variables in the lab frame 
\begin{eqnarray}
\label{CM-vectors-back}
&&\hspace{-1cm}p^{\,' 0}_{k^{\, '}}=\gamma \, \left( p^{\,' 0\; cm}_{k^{\, '}} + \vec{\beta} \cdot 
\vec{p}^{\,' \; cm}_{k^{\, '}} \right) \; , \\
&&\hspace{-1cm}\vec{p}^{\,'}_{k^{\, '}}=\vec{p}^{\,' \; cm}_{k^{\, '}} +\frac{1}{\beta^2} \,
   (\gamma -1) \, (\vec{p}^{\,' \; cm}_{k^{\, '}} \cdot 
\vec{\beta}) \, \vec{\beta} + \gamma \, \vec{\beta} \, p^{\,' 0 \;
   cm}_{k^{\, '}} 
\; , \\
&&\hspace{-1cm}p^{\, ' 0}_{l^{\, '}}=\gamma \, \left( p^{\,' 0\; cm}_{l^{\, '}} + \vec{\beta} \cdot
\vec{p}^{\,' \; cm}_{l^{\, '}} \right) \; , \\
&&\hspace{-1cm}\vec{p}^{\,'}_{l^{\, '}}=\vec{p}^{\,' \; cm}_{l^{\, '}} +\frac{1}{\beta^2} \, (\gamma -1) \,
   (\vec{p}^{\,' \; cm}_{l^{\, '}} \cdot 
\vec{\beta}) \, \vec{\beta} + \gamma \, \vec{\beta} \, p^{\,' 0 \;
   cm}_{l^{\, '}} 
\; .
\end{eqnarray}
At this point we have the momenta of the nuclear fusion products in
the $k^{\, '}l^{\, '}$-system in the lab frame again. We note that the
masses of the products are typically different. However,
total energy and momentum are conserved.

Since electrons are relativistic and to derive simple scaling laws for
nuclear fusion efficiency at a later point in this paper three
notations are helpful. The three notation are obtained by performing
the integration over $\vec p^{\,'}_{l^{\, '}}$ in (\ref{kl-system}). We
obtain $\vec p^{\,'}_{l^{\, '}}=\vec p_k + \vec p_l - \vec p^{\,'}_{k^{\,
'}}$. Making use of the center of mass frame we find
\begin{eqnarray}
\label{delta-function-subst}
&&\hspace{0cm}\frac{\delta \left( p^{0 \, cm}_k+p^{0 \, cm}_l - p^{\,'
   0\,
   cm}_{k^{\, '}}-p^{\,' 0 \, cm}_{l^{\, '}} \right)}{p^{\,' 0 \, cm}_{k^{\,
   '}} p^{\,' 0 \, cm}_{l^{\, '}}} \\
&&=\frac{\delta \left( |\vec{p}^{\,' \;cm}_{k^{\, '}}|-\frac{{\cal
   F}_{kl}}{\sqrt{s}} \right)}{|\vec{p}^{\,' \;cm}_{k^{\, '}}| \sqrt{s}} \,
   , \nonumber
\end{eqnarray}
where the quantity ${\cal F}_{kl}$ is given by 
\begin{eqnarray}
  \label{inv-flux}
  &&{\cal F}_{kl}=\sqrt{\left( p_k \cdot p_l \right)^2-m^2_k m^2_l c^4} \, .
\end{eqnarray}
In the velocity space this leads to
\begin{eqnarray}
\label{transport-eqn-fusion1}
&&\frac{\partial f_k}{\partial t} + \vec v_k \cdot \frac{\partial
f_k}{\partial \vec x_k} + q_k \left( \vec E + \vec v_k \times \vec B \right) \cdot
   \frac{\partial f_k}{\partial {\vec p_k}} \\
&=&\sum_{l} \int d^3p_l \, v^{kl}_{rel} \, 
   \int d\Omega_{\psi} \; \sigma^{kl}_C \left( s, \psi \right) \, \left(  f_{l^{\, '}}  f_{k^{\, '}} -  f_l f_k
   \right) \nonumber \\
&&- \sum_{l} \int d^3p_l \, v^{kl}_{rel} \, 
\int d\Omega_{\psi} \; \sigma^{kl}_R \left( s, \psi \right) \, f_k \, 
   f_l  \, ,  \nonumber
\end{eqnarray}
where
\begin{eqnarray}
v^{kl}_{rel} &=& \frac{c \, {\cal F}_{kl}}{p^0_k \, p^0_l} =\sqrt{\left| \vec v_k - \vec v_l \right|^2 -
                 \frac{1}{c^2} \left( \vec v_k \times \vec v_l \right)^2} \, , \\
\vec v_k &=& \frac{c \vec p_k}{\sqrt{m^2_k c^2 + \vec p^2_k}} \, .
\end{eqnarray}
We finally find
\begin{eqnarray}
\label{transport-eqn-fusion2}
&&\frac{\partial f_k}{\partial t} + \vec v_k \cdot \frac{\partial
f_k}{\partial \vec x_k} +\frac{q_k}{m_k} \left( \vec E + \vec v_k \times \vec B \right) \cdot
   \frac{\partial f_k}{\partial {\vec v_k}} \\
&=&\sum_{l} \int d^3v_l \, v^{kl}_{rel} \, \int d\Omega_{\psi} \; \sigma^{kl}_C \left( s, \psi \right) \,
   \left(  f_{l^{\, '}}  f_{k^{\, '}} -  f_l f_k
   \right) \nonumber \\
&&- \sum_{l} \int d^3v_l \, v^{kl}_{rel} \, \int d\Omega_{\psi} \; \sigma^{kl}_R \left( s, \psi \right) \, f_k \,
   f_l  \, ,  \nonumber
\end{eqnarray}
where fuel breeding is excluded.

In section \ref{nano_accelerator} we parametrically discuss the
integrated nanoscopic accelerator concept. It relies on the assumption
that the driver laser is capable of removing sufficiently many
electrons from the nano-rods in the micro-reactor in a predictable
way. We will not investigate the details in this paper and leave it
for future numerical analysis.

\section{The integrated nano-accelerator} \label{nano_accelerator}
We assume that energy transfer to fuel constituents takes place with
the help of the electromagnetic fields of the driver laser. This
energy delivery system has a chance to be efficient and fast for
almost all nuclear fusion fuels.

The convertor concept discussed in the present paper consists of a
laser-powered integrated nano-structured accelerator.
For reasons of efficiency it consists exclusively of very small
nuclear fuel based nano-structures that allow Coulomb explosions.

The integrated nano-accelerator is assumed to be an efficient design
for the generation of large ionic currents at low ion energies. It is
powered by ultra-short ultra-high energy laser pulses in the UV
to the VUV wavelength range. The integrated accelerator is expected to
absorb the driver laser energy almost completely avoiding parametric
optical instabilities.

Nano-structures can be efficient laser energy convertors into ionic motion
if the electrons in the nano-structures can be over-heated by the
driver laser. Over-heated electrons are those that cannot be
recaptured by the ionized nano-structures within the time window the
Coulomb explosions take. Hence, positive ions are exposed to their own
electric space-charge field for long enough and Coulomb explode,
leaving behind after some time, a nearly homogeneous
ionic distribution in configuration space and a non-thermal one in the
momentum space that can be engineered such that it is peaked at the
resonances of the provided nuclear fuel mix.

Specifically, we consider cylindrical nano-rods that form the
embedded nano-accelerator as sketched in Fig. \ref{absorber}, which is
composed of rigid fuel constituents $l$ into which lighter fuel
constituents $k$ are embedded. We assume $e_l/m_l \ll e_k/m_k$ for the
effective charges and fuel masses involved. Between accelerator nano-rods
low Gamov energy fuel ions can be places in form of foams. Since we
propose UV to VUV driver laser wavelengths we assume that the laser is
still capable of propagating through the proposed convertor in a
stable and predictable way.

The VUV laser driver ionizes the nano-rods partially. The ionizing
electrons occupy the space inside and between the nano-rods in such a
way that individual nano-rods are partially shielded from each other.
Still they provide an ion accelerating electric field strong enough to
obtain the required relative energies between the fuel constituents
for the provided fusion resonances of the fuel mix.

For reasons of simplicity we make the following assumption for the
electric field of a single nano-rod composed of the high density fuel
constituent $l$
\begin{eqnarray}
\label{electric-field}
&&E_r \left( r_k \right) = \left\{
\begin{array}{ll}
C_l \, r_k  \, , & 0 \le r_k < R \\
0  \, , & r_k \ge R \\
\end{array}
\right. \, , \quad C_l= \frac{e \, n_l}{2 \, \epsilon_0} \, ,
\end{eqnarray}
where the nanoscopic field $E_r$ is radial and $n_l$ represents the
average positive charge density inside the rods. The parameter $r_k$
is the radial position of an ion of sort $k$ inside the nano-rod
composed of ions of sort $l$ and $R$ is the nano-rod radius.

Since collisions and fusion reactions are rare on $\text{fs}$ time
scales, which the postulated Coulomb explosions require, we neglect
the collision and fusion operators in (\ref{transport-eqn-fusion2}) during the
Coulomb explosion phase. In addition, we assume that the light ions
only expand radially, while the heavy ones stand still. Hence, for $r_k
\le R$ the acceleration of the $k$-ions is approximated by the
following radial Vlasov equation
\begin{eqnarray}
\label{light-ion-vlasov1}
&&\hspace{-1cm}\left( \partial_t + v_k \partial_{r_k} + \frac{e_k}{m_k} \, C_l r_{k} \,
\partial_{v_k} \right) \, \left( r_k v_k f_k \right) \left( r_k, v_k, t \right)
= 0 \, ,
\end{eqnarray}
where $f_k$ now is a distribution funstion in polar coordinates.
The above approximation is justified for the $\ce{p^{11}B}$ fuel, for
example. For $r_k > R$ the light ions undergo further acceleration. Also
the heavy ions $l$ are ultimately subject to acceleration. However, for
simplicity we neglect secondary forces on all fuel ions $kl$. This implies
for the light ions of sort $k$ for $r_k>R$
\begin{eqnarray}
\label{light-ion-vlasov2}
&&\left( \partial_t + v_k \partial_{r_k} \right) \, \left( r_k v_k f_k \right) \left( r_k, v_k, t \right)
   = 0
\end{eqnarray}
until they collide or fuse.

According to (\ref{light-ion-vlasov1}) the light ions fulfill the
following equations of motion during the Coulomb explosion phase
\begin{eqnarray}
\label{eqns1}
&&\frac{dr_k}{dt} = v_k \, , \quad \frac{dv_k}{dt} = \frac{e_k}{m_k} \, C_l \,r_k \, ,
\end{eqnarray}
while the solution of (\ref{light-ion-vlasov1}) is
\begin{eqnarray}
\label{eqns3}
&&\left( r_k v_k f_k \right) \left( r_k, v_k, t \right) = \left( r_{k0} v_{k0} f_k \right)
   \left( r_{k0}, v_{k0}, 0 \right) \, ,
\end{eqnarray}
where due to  (\ref{eqns1}) we have
\begin{eqnarray}
\label{solution_inv}
&&\hspace{-1.2cm}\left(
\begin{array}{c}
 r_{k0} \\
 v_{k0} \\
\end{array}
\right) \\
&&\hspace{-1.0cm}=
\left(
\begin{array}{cc}
\cosh \left( \sqrt{\frac{e_kC_l}{m_k}} t \right) &
-\frac{1}{\sqrt{\frac{e_kC_l}{m_k}}}
\, \sinh \left( \sqrt{\frac{e_kC_l}{m_k}} t \right) \\
-\sqrt{\frac{e_kC_l}{m_k}} \, \sinh \left( \sqrt{\frac{e_kC_l}{m_k}} t \right)
& \cosh \left( \sqrt{\frac{e_kC_l}{m_k}} t \right) \\
\end{array}
\right) \,
\left(
\begin{array}{c}
r_k \\
v_k \\
\end{array}
\right) \, . \nonumber
\end{eqnarray}
The parameter $r_{k0} \le R$ is the initial radial position of the
light ions and $v_{k0}>0$ their initial radial velocity. The
parameters $r_k$ and $v_k$ are the radial position and velocity
at times $t>0$.

\begin{figure}
  \begin{center}
  \includegraphics[width=60mm]{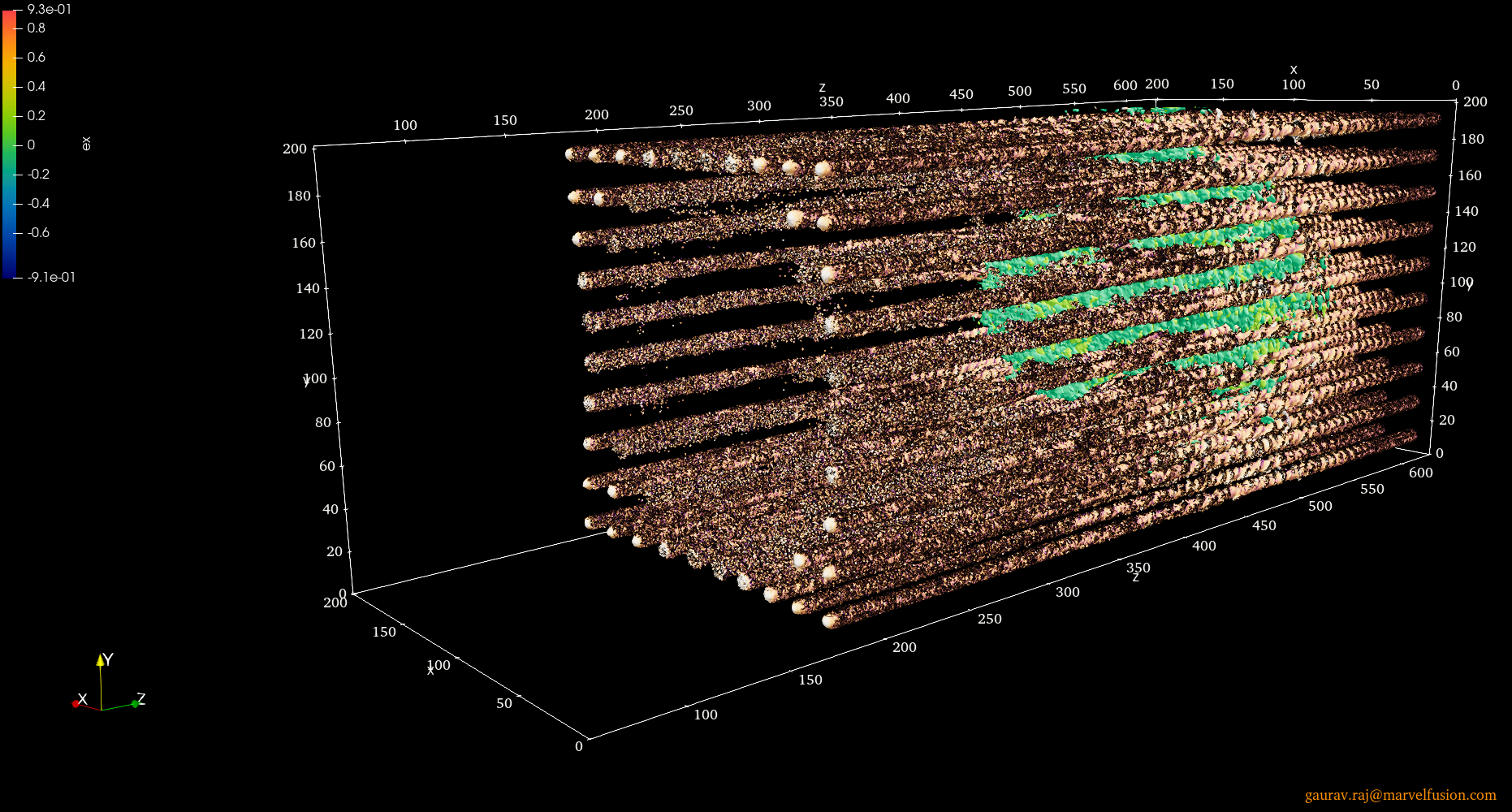}\\
  \includegraphics[width=60mm]{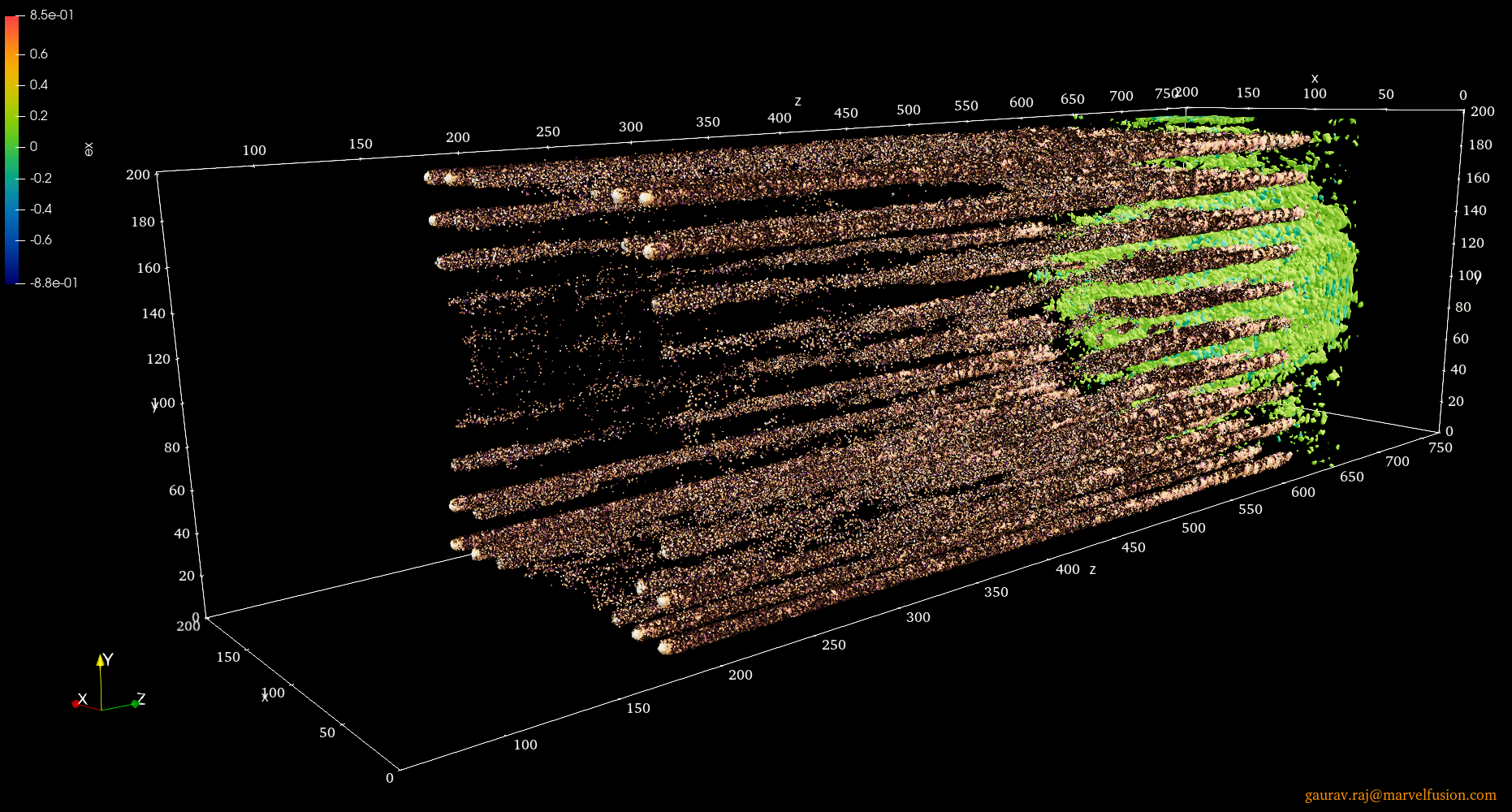}
\end{center}
\caption{\label{coulomb1} The laser impinges on the nano-reactor from the 
left. Two time frames are shown. The box size is $8 \, \mu \text{m}
\times 8 \, \mu \text{m} \times 32 \, \mu \text{m}$. The nano-rods are 
$30 \, \text{nm}$ wide and about $800 \, \text{nm}$ apart. The laser 
wavelength is $250 \, \text{nm}$. The laser is circularly 
polarized. The laser driver has nearly constant loss of energy per 
unit of propagation length. The laser does not self-focus. It captures 
electrons forming a magnetic field that separates electron current and 
return currents. The plot shows that the plasma starts to pinch. The 
simulation is the same as the one in Fig. \ref{absorber}.}
\end{figure}

\begin{figure}
  \begin{center}
    \includegraphics[width=50mm]{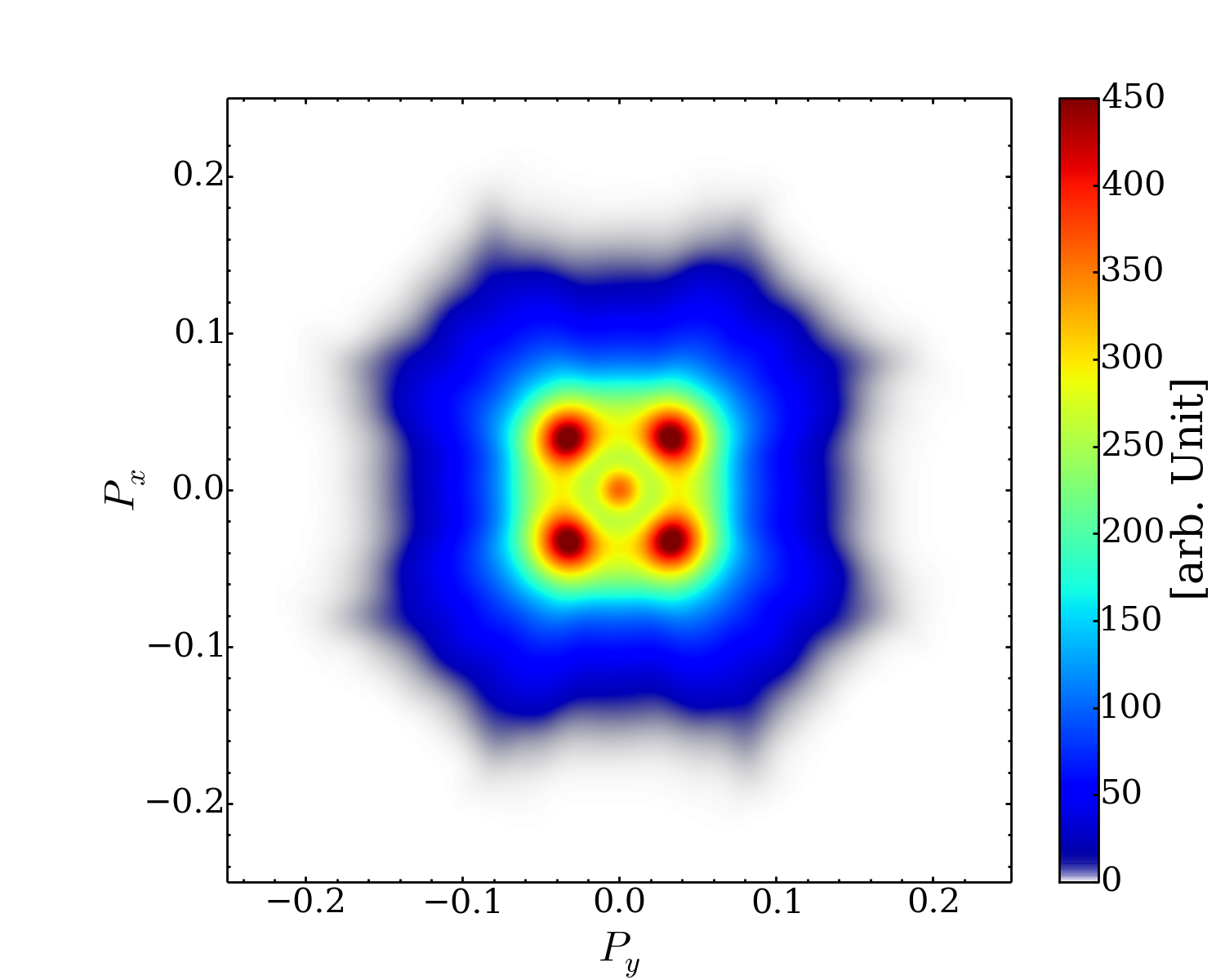}\\
    \includegraphics[width=50mm]{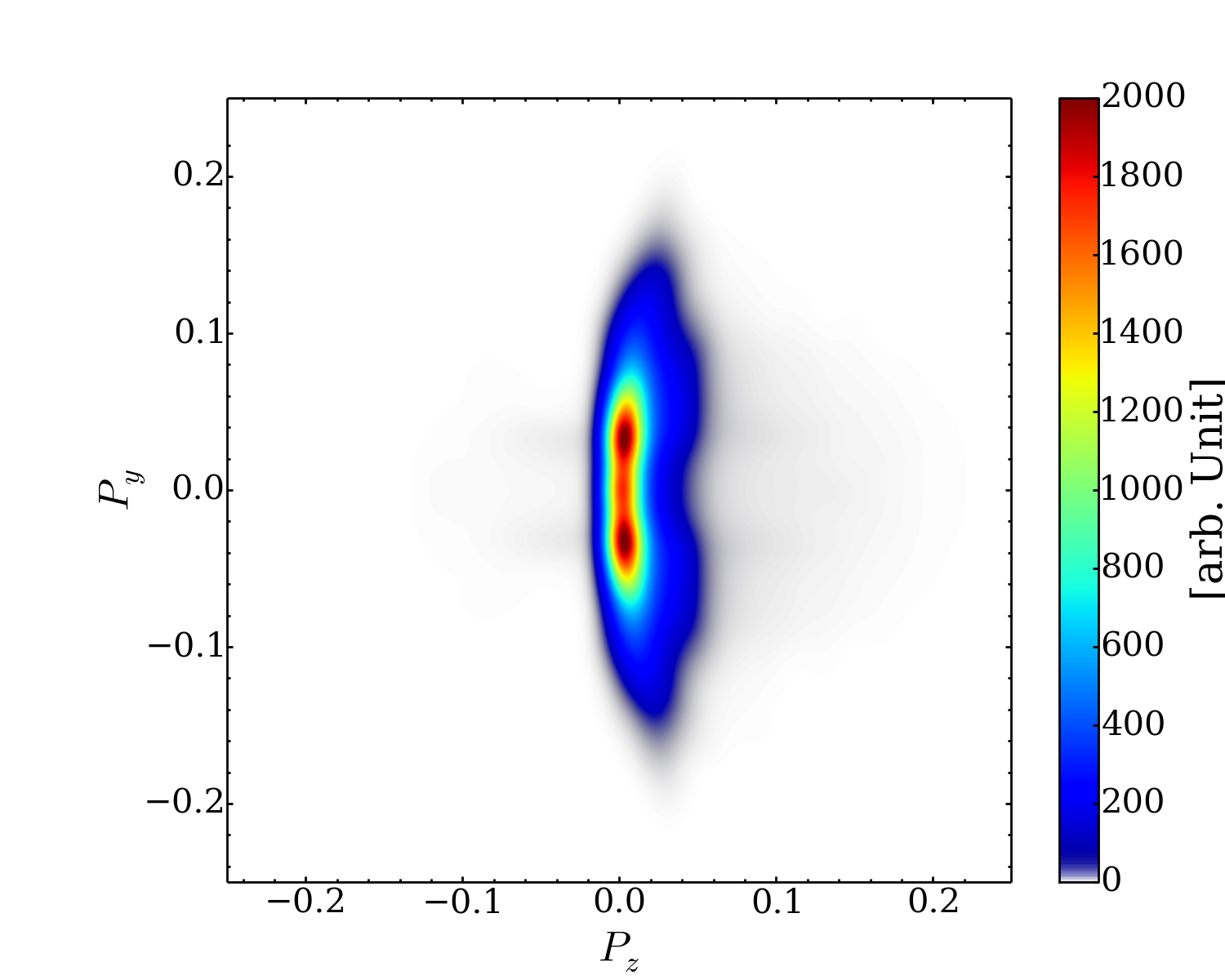}
  \caption{\label{proton_pxpy_distr} Proton momentum distributions 
    integrated over the configuration space after the laser 
    pulse has exited the nano-structures. The proton momenta are 
    normalized to $m_{\text{p}}c$. The simulation has lateral periodic 
    boundaries, while the laser pulse is a central disk with $10 \, 
    \text{fs}$ pulse length and $\lambda=250 \, \text{nm}$. The 
    polarization is circular. The simulation is the same as the one in 
  Fig. \ref{absorber}.}
  \end{center}
\end{figure}

\begin{figure}
  \begin{center}
    \includegraphics[width=50mm]{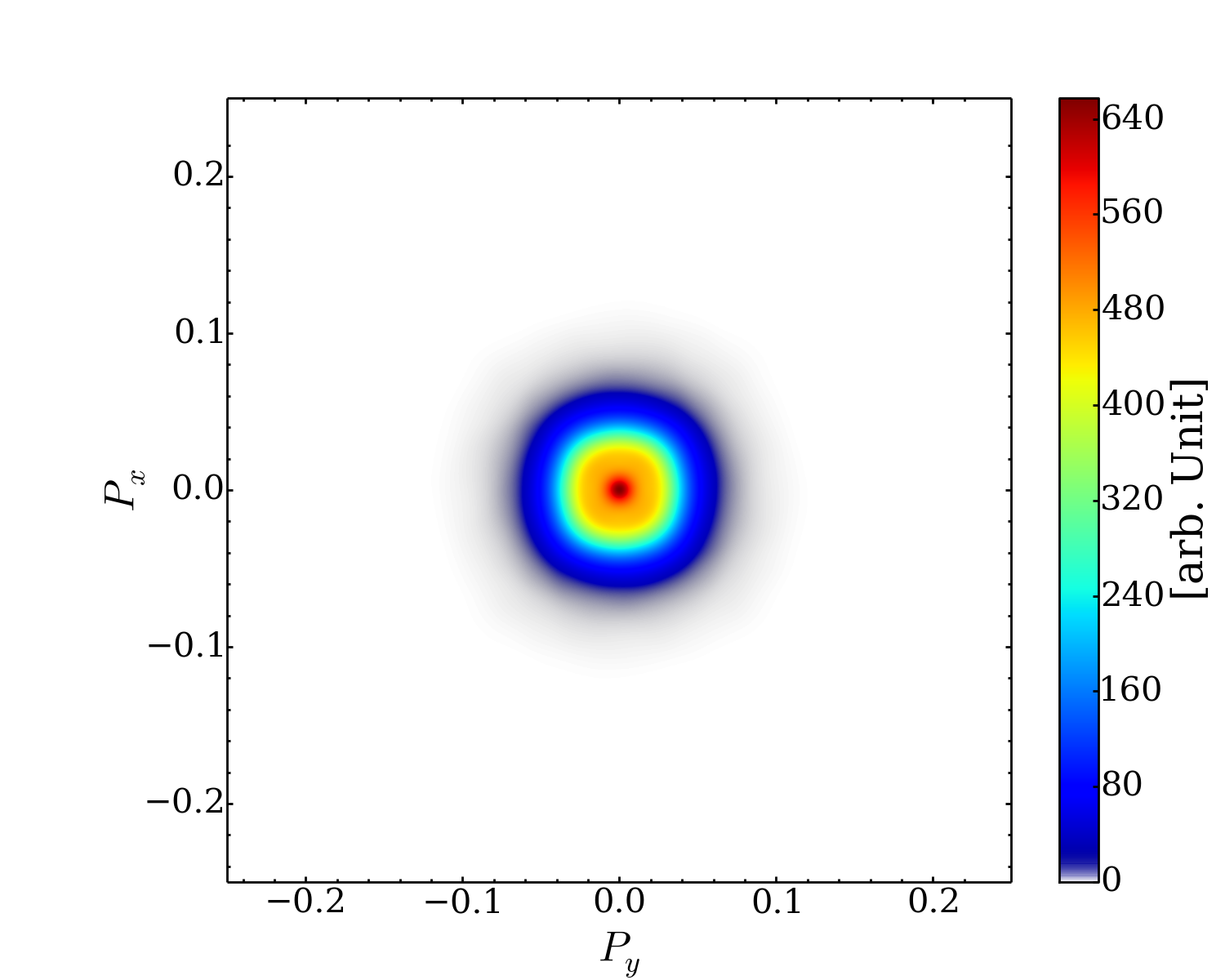}\\
    \includegraphics[width=50mm]{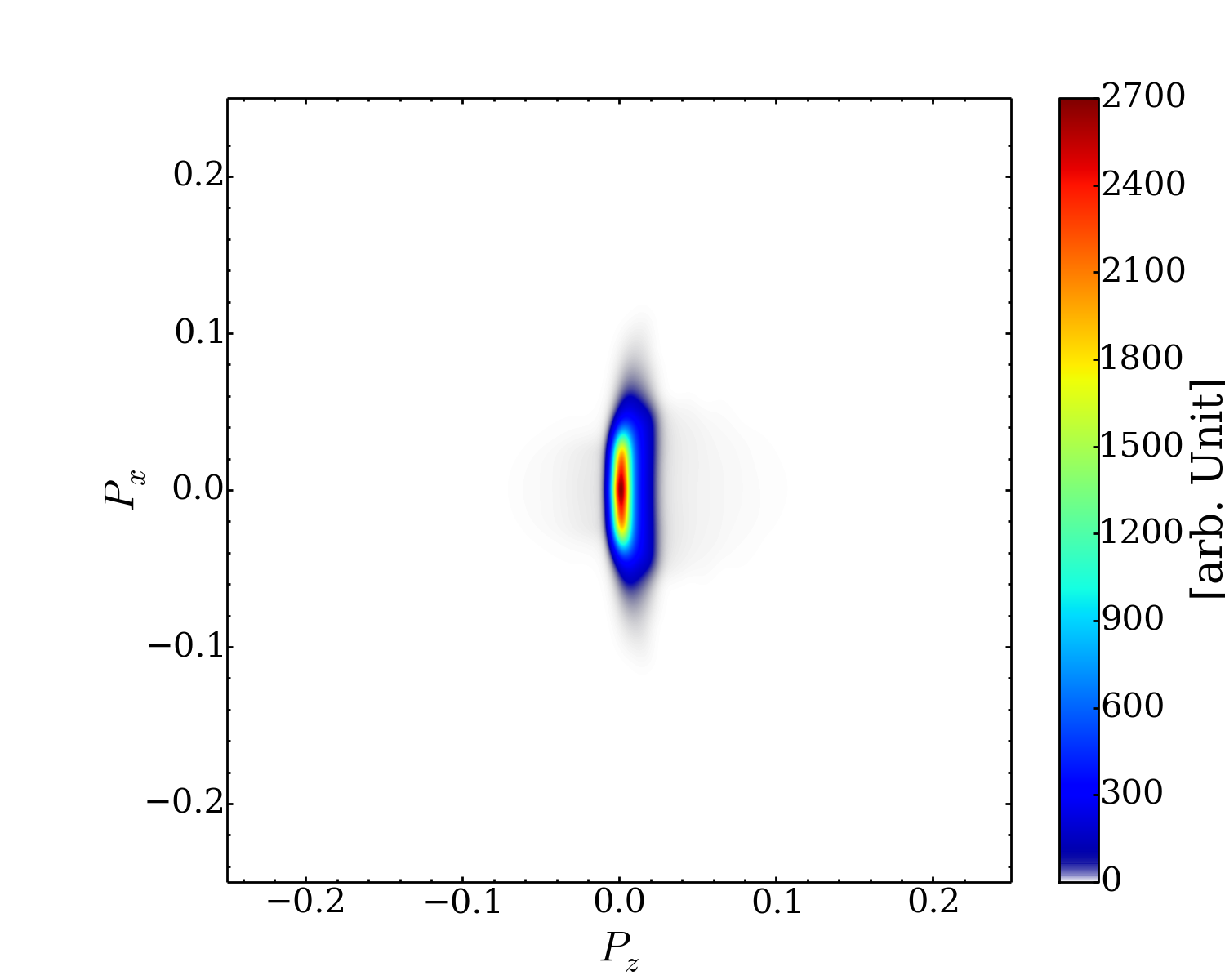}
  \caption{\label{boron_pxpy_distr} Boron momentum distributions 
    integrated over the configuration space after the laser 
    pulse has exited the nano-structures. The boron momenta are 
    normalized to $m_{\text{B}}c$. The simulation has lateral periodic 
    boundaries, while the laser pulse is a central disk with $10 \, 
    \text{fs}$ pulse length and $\lambda=250 \, \text{nm}$. The 
    polarization is circular. The simulation is the same as the one in 
  Fig. \ref{absorber}.}
  \end{center}
\end{figure}

To estimate the energy distribution of the light ions we consider a
layer $s$ of the latter with initial radial position $0 < r^s_k(0) \le R$ and
a radial velocity distribution given by
\begin{eqnarray}
\label{distribution}
&&\left( r_{k0} v_{k0} f_k \right) \left( r_{k0}, v_{k0},
   0 \right) \\
&&= \frac{N^s_k}{4\pi^2} \, \delta
   \left( r_{k0} - r^s_k(0) \right) \, \delta \left( v_{k0} - v^s_k(0)
   \right) \, , \nonumber
\end{eqnarray}
where $N^s_k$ is the initial number of particles $k$ at the radial position
$r^s_k(0)$. Plugging $\ce{r}_{k0}$ and $v_{k0}$ given by (\ref{solution_inv}) into
(\ref{distribution}) assuming $v^s_K(0) =0$ gives for  $t \le t^s_k$
\begin{eqnarray}
\label{velocity_ring1}
&&\hspace{-1.0cm}r_k \, v_k \, f^s_k \left( r_k, v_k, t \right) = \frac{N^s_k}{4\pi^2}  \, \delta
\left( r_k - r^s_k(t) \right) \, \delta \left( v_k - g^s_k(t) \right) \, ,
\end{eqnarray}
where
\begin{eqnarray}
  \label{texit}
&&t^s_k = \sqrt{\frac{m_k}{e_kC_l}} \, \cosh^{-1} \left( \frac{R}{r^s_k(0)}
   \right) \, , \\
&&r^s_k(t) = r^s_k(0) \, \cosh \left( \sqrt{\frac{e_kC_l}{m_k}} t \right) \, ,\\
&&g^s_k(t) = \sqrt{\frac{e_kC_l}{m_k}} \, r^s_k(0) \, \sinh \left(
   \sqrt{\frac{e_kC_l}{m_k}} t \right) \, .
\end{eqnarray}
After rapid acceleration during the Coulomb explosion phase the light
ion layer is assumed to move on ballistically. At $t^s_k$ we obtain
\begin{eqnarray}
\label{velocity_ring_exit}
&&\hspace{-1.0cm}r_k \, v_k \, f^s_k \left( r_k , v_k, t^s_k \right)
   = \frac{N^s_k}{4\pi^2}  \, \delta \left( r_k - r^s_k(t^s_k) \right) \,
\delta \left( v_k - g^{s}_k(t^s_k) \right) \, ,
\end{eqnarray}
where
\begin{eqnarray}
\label{exit_position}
&&r^{s}_k(t^s_k) = R \, , \\
\label{exit_velocity}
&&g^{s}_{k}(t^s_k) = \sqrt{\frac{e_kC_l}{m_k}} \, \sqrt{R^2 - \left(
   r^s_k(0) \right)^2} \, .
\end{eqnarray}
It holds that
\begin{eqnarray}
\label{norm}
&&\hspace{-1cm} ( 2\pi)^2 \, \int^{\infty}_0 dv_k \, v_k \int^{\infty}_0 dr_k \, r_k  
   \,  f^s_k \left( r_k, v_k, t \right) = N^s_k \, .
\end{eqnarray}
Simulations confirm that an approximately homogeneous and 
isotropical light ion distribution in configuration space is obtained 
after the interaction with the laser pulse, while a peaked non-thermal 
light ion distribution in momentum space remains. Figure
\ref{coulomb1} shows a simulation of the laser impinging on the
nano-structures, which extend beyond the interaction volume with the
laser. Figure \ref{proton_pxpy_distr} shows the proton momentum and
Fig. \ref{boron_pxpy_distr} the boron momentum distribution obtained
from the same simulation.
The distribution function $f_p$ of the protons is peaked 
at small momenta mainly due to rods not interacting with the laser 
pulse and at $|\vec p_{\text{p}}| \approx 0.03 \, m_{\text{p}} c$ due 
to nano-acceleration and periodic boundaries used in configuration 
space. There are also many protons at $|\vec p_{\text{p}}| >0.05 \, m_{\text{p}}c$. 
\begin{figure}
  \begin{center}
  \includegraphics[width=60mm]{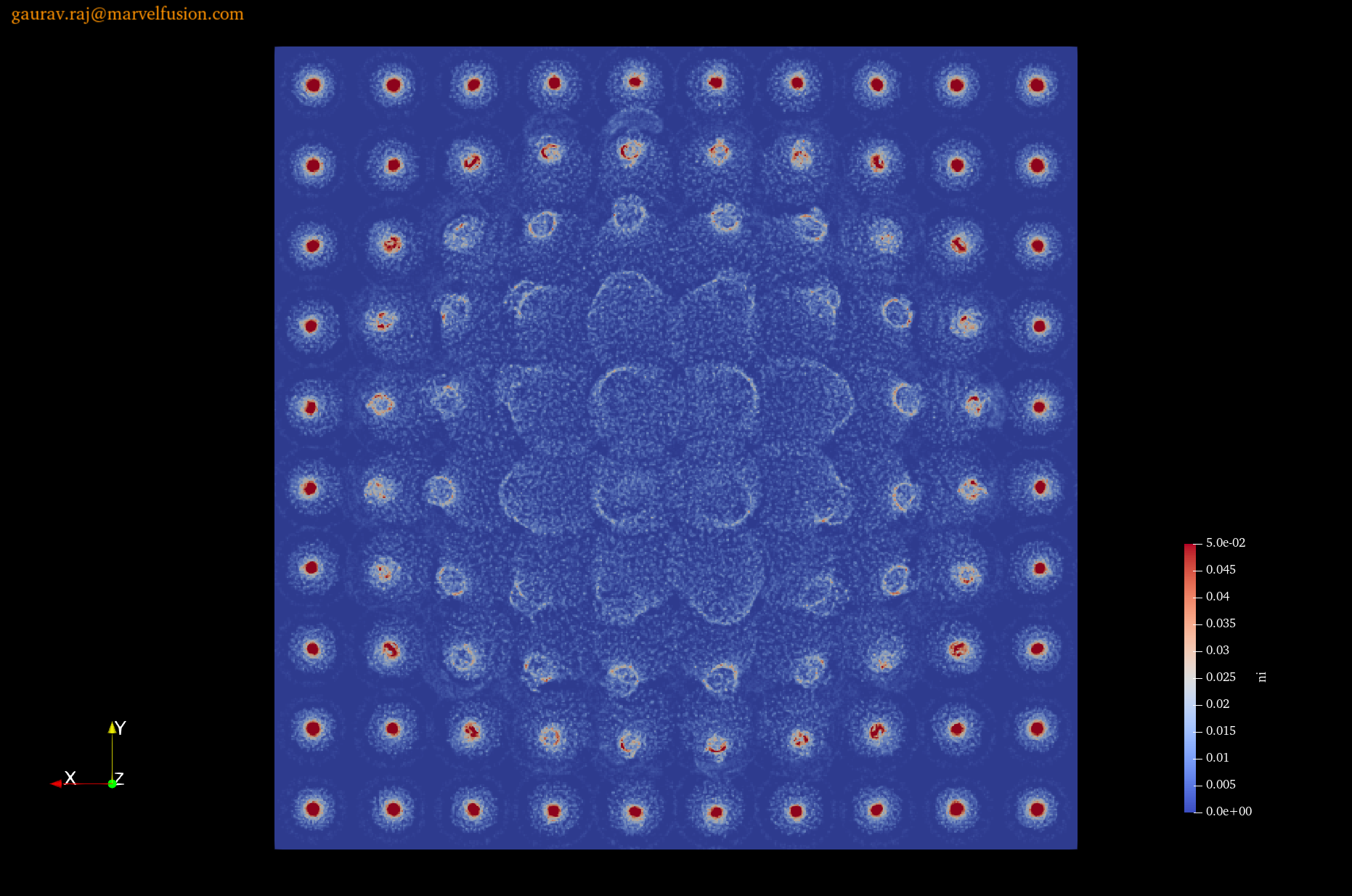}
\end{center}
\caption{\label{coulomb2} Coulomb explosions in the body of the laser 
pulse. The explosion front propagates with almost the speed of 
light. The simulation is the same as the one in Fig. \ref{absorber}.}
\end{figure}

\section{Conversion fraction} \label{burn_fraction}
To derive a relation for the burn fraction we make use of  
(\ref{transport-eqn-fusion2}). Next, we need parametrizations of the
ionic distribution functions. We assume that the ions of sort $k$ are
mobile and make the ansatz
\begin{eqnarray}
\label{distr_approx_k}
&&\hspace{-0.8cm} f_k \left( \vec r_k, \vec v_k, t \right) \\
&&\approx
   \sum_{i, s} \alpha^s_k \, N^i_{k}(t) \, \delta^3 
   \left( \vec r_k -  \vec r^{\, s,i}_{k}\left( t \right) \right) \, \delta^3 
   \left( \vec v_k -  \vec g^{\, s,i}_{k}(t) \right)\, . \nonumber
\end{eqnarray}
We further assume that the ions of sort $l$ are immobile. We make the
ansatz
\begin{eqnarray}
\label{distr_approx_l}
&&\hspace{-0.8cm}f_l \left( \vec r_l, \vec v_l, t \right) \\
&&\approx \frac{N}{V} \, \sum_i N^i_{k}(t) \, \delta^3 \left( \vec
v_{l} \right) \, . \nonumber
\end{eqnarray}
Summation over all fluid elements $i$ implies
\begin{eqnarray}
\label{N_k}
&&N_k=\sum_i N^i_{k} \, , \\
\label{n_l}
&&n_l(t) = n_k(t) = \frac{N}{V} \, N_k(t) \, .
\end{eqnarray}
In addition we have
\begin{eqnarray}
\label{alpha_k}
&&\alpha^s_k =\frac{2r^s_k(0) \, \Delta r_k}{R^2} \, , \\
\label{sum_alpha_k}
&&\sum_s \alpha^s_k = 1 \, .
\end{eqnarray}
The parameter $N$ is the number of nano-rods in the reactor, that can
be reached by the accelerated fuel constituent $k$, and $\Delta r_k$
is the thickness of the layer $s_k$ of the fuel constituent $k$ in a
single nano-rod. We assume that all initial velocity directions of
constituent $k$ are uniformly distributed while their absolute
velocity values are given by (\ref{exit_velocity}). The initial
positions are determined by the positions of the nano-rods.
In addition, each velocity group of constituent $k$ has its own 
density group. All density groups of constituent $k$ add up to the
total density.

Intra-ionic collision do not drain much energy from the ionic
subsystem. They predominantly redistribute phase-space without much
energy loss. Hence, we neglect them. Electron - ion collsions,
however, do not change the direction of the ions very much, but are
potentially capable of draining a lot of energy from the
ions. Assuming an adequate shape of the electronic distribution
function $f_e$ the collision integral in (\ref{transport-eqn-fusion2}) 
leads to resistivities. We do not discuss further details here.

With the help of the zero and first order velocity moments of
(\ref{transport-eqn-fusion2}), neglecting the electromagnetic fields
in (\ref{transport-eqn-fusion2}), making use of (\ref{distr_approx_k})
- (\ref{sum_alpha_k}) and our assumption about the electronic
distribution function we obtain for the ions of sort $k$ of a single rod
\begin{eqnarray}
\label{rate1}
&&\hspace{-0.5cm}\frac{dN^i_{k}(t)}{dt} \approx - N^{i}_{k}(t) \, N_k(t) \,
   \frac{N}{V} \, \sum_s \alpha^s_k \, g^{s,i}_k(t) \, \sigma^{kl}_R \left(
   g^{s,i}_k(t) \right) \, ,
\end{eqnarray}
where
\begin{eqnarray}
\label{rate2} 
&&\hspace{-0.5cm}\frac{d \vec g^{\, s,i}_k(t) }{dt} 
\approx - \nu^{s,i}_{ke}\left( \vec g^{\, s,i}_k(t) \right) \,
  \vec g^{\, s,i}_k(t)
\end{eqnarray}
and
\begin{eqnarray}
\label{rate3}
&&\hspace{-0.5cm}\frac{d\vec r^{\, s,i}_k(t)}{dt} = \vec g^{\, s,i}_k(t) \, .
\end{eqnarray}
The $\nu^{s,i}_{ke}$ are the resistivities approximately given by
\begin{eqnarray}
\label{collision-frequencies}
&&\nu^{s,i}_{ke} \left( \vec g^{\, s,i}_k (t) \right) \approx \frac{e^2_k e^2_e \,
\frac{N}{V} \, N^i_k}{4 \pi \epsilon^2_0 m^2_{ke} \,
\left| \vec g^{\, s,i}_k (t) - \vec v_e(t) \right|^3} \, \ln \Lambda_{ke} \, ,
\end{eqnarray}
where $\vec v_e$ is the electron velocity and $\ln \Lambda_{ke}$ is the
Coulomb logarithm.

Equations (\ref{rate1}) - (\ref{rate3}) are quasi-element equations
for fuel ions. As they are they cannot be solved. However, assuming
that all nano-rods are the same and that there is rotational and
translational invariance of the system the index $i$ can be dropped
and we obtain the much simpler system of equations
\begin{eqnarray}
\label{rate1n}
&&\hspace{-0.5cm}\frac{dN_{k}(t)}{dt} \approx - N^2_{k}(t) \,
   \frac{N}{V} \, \sum_s \alpha^s_k \, g^{s}_k(t) \, \sigma^{kl}_R \left(
   g^{s}_k(t) \right) \, ,
\end{eqnarray}
where
\begin{eqnarray}
\label{rate2n} 
&&\hspace{-0.5cm}\frac{d \vec g^{\, s}_k(t) }{dt} 
\approx - \nu^{s}_{ke}\left( \vec g^{\, s}_k(t) \right) \,
  \vec g^{\, s}_k(t)
\end{eqnarray}
and
\begin{eqnarray}
\label{rate3n}
&&\hspace{-0.5cm}\frac{d\vec r^{\, s}_k(t)}{dt} = \vec g^{\, s}_k(t) \, .
\end{eqnarray}
Mean field and binary level radiation loss effects are neglected in
(\ref{rate2n}) - (\ref{rate3n}). 

\begin{figure}[ht]
\begin{center}
  \includegraphics[width=70mm]{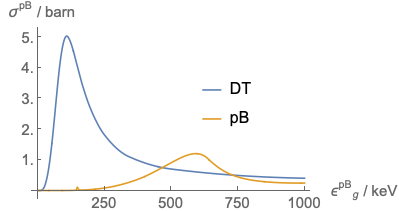}  
\end{center}
\caption{\label{pB_cross_section_paper}  Comparison between the 
cross sections of $\ce{pB}$ and $\ce{DT}$ as quoted in reference 
\cite{nevins2000thermonuclear}.}
\end{figure}

Equation (\ref{rate1n}) decouples from (\ref{rate2n}) and
(\ref{rate3n}). It can be solved. We find
\begin{eqnarray}
\label{proton_evolution}
&&\hspace{-1cm}\Delta N_k \approx \frac{N_k}{1+ \frac{N}{V} \, N_k \,
\int^{\infty}_{t^s_k} dt \sum_s \alpha^s_k \, g^{s}_k (t) \, \sigma^{kl}_R \left(
g^{\, s}_k(t) \right)}  \, ,
\end{eqnarray}
The parameter $N_k$ is a free parameter. The conversion fraction is
\begin{eqnarray}
\label{conversion-fraction1}
\eta^{kl}&=&1-\frac{\Delta N_k}{N_k} \\
&\approx& \frac{\frac{N}{V} \, N_k \, \sum_s \alpha^s_k \,
\int^{\infty}_{t^s_k} dt \, g^{s}_k (t) \, \sigma^{kl}_R \left(
    g^{\, s}_k(t) \right)}{1+\frac{N}{V} \, N_k \, \sum_s \alpha^s_k \,
    \int^{\infty}_{t^s_k} dt \, g^{s}_k (t) \, \sigma^{kl}_R \left(
    g^{\, s}_k(t) \right)} \, . \nonumber
\end{eqnarray}
To obtain the velocity (\ref{rate2n}) has to solved. We find
\begin{eqnarray}
&&\vec g^s_k(t) = \vec g^s_k\left( t^s_k \right) \, e^{-\nu^s_{ke} \,
   \left( t - t^s_k \right)} \, , \quad t \gg t^s_k \, .
\end{eqnarray}
We assume that the fusion cross section is constant within a given
velocity range
\begin{eqnarray}
\label{sigma_approx}
&&\sigma^{kl}_R \left( g^{s}_k(t) \right) =
   \left\{
   \begin{array}{ll}
     \sigma^{kl}_0 \, , & \sqrt{\frac{2\epsilon^{kl}_1}{m_k}} \le 
     g^{s}_k(t) \le \sqrt{\frac{2\epsilon^{kl}_2}{m_k}} \\
     0 \, , & \text{else} \\
   \end{array}
   \right. , 
\end{eqnarray}
where $\sigma^{kl}_0$ is the smallest value within the velocity range.
A comparison of the cross sections for $DT$ and $pB$ is given in
Fig. \ref{pB_cross_section_paper}. It holds $\sigma^{DT,pB}_0 \approx
10^{-28} \, \text{m}^2$ for $0 < g^s_k(t) < 2 \cdot 10^7 \,
\text{ms}^{-1}$.

There are many interesting fuel cycles from which $\sigma^{kl}_0$ can
be obtained. The cross sections can be parametrized as
\begin{eqnarray}
&&\sigma^{kl}_R \left( g^{\, s}_k(t) \right) \approx
\frac{S_{kl}\left( \frac{\epsilon^{kl}_{G}}{\beta^{kl}_{av}}
\right)}{\epsilon^{kl}_{G}} \, \beta^{kl}_{av} \,
e^{-\sqrt{\beta^{kl}_{av}}} \, ,
\end{eqnarray}
where 
\begin{eqnarray}
&&\beta^{kl}_{av} = \frac{\epsilon^{kl}_{G}}{\epsilon^{kl}_{av}}  \, , \\
&&\epsilon^{kl}_{av} = \frac{1}{2} \, m_{kl} \, \left( g^{av}_k \right)^2 \, , \\
&&\epsilon^{kl}_{G} = 2 \, \left( \frac{\pi e^2 Z_kZ_l}{4 \pi 
   \epsilon_0 \hbar c} \right)^2 \, m_{kl} c^2 \, , \\
&&m_{kl}=\frac{m_k \, m_l}{m_k + m_l} \, . 
\end{eqnarray}
The most relevant neutronic fuel cycles are given in table
\ref{neutronic-cycles}. They have small Gamov energies
$\epsilon^{kl}_{G}$ and large $S_{kl}$.
\begin{table}[ht]
\begin{center}
{\small
\begin{tabular}{|l|c|c|c|}
    \hline
    standard fuels & $\epsilon^{kl}_f$ MeV & $S_{kl}$ keV barn & $\sqrt{\epsilon^{kl}_{G}} \; \sqrt{\text{keV}}$
    \\
    \hline
    $\ce{D}+\ce{T} \rightarrow \ce{^4He} +\ce{n}$ & $17.59$ & $1.2 \cdot 10^4$ & $34.38$\\
    $\ce{D}+\ce{D} \rightarrow \ce{T} +\ce{p}$ & $4.04$ & $56.0$ &$31.4$ \\
    $\ce{D}+\ce{D} \rightarrow \ce{^3He} +\ce{n}$ & $3.27$ & $54.0$
                            &$31.4$ \\
    $\ce{D}+\ce{D} \rightarrow \ce{^4He} +\ce{\gamma}$ & $23.85$ &
                                                                   $4.3 
                                                                   \cdot 
                                                                   10^{-3}$
                            &$31.4$ \\
    $\ce{T}+\ce{T} \rightarrow \ce{^4He} +2\ce{n}$ & $11.33$ &
                                                                   $138.0$
                            &$38.45$ \\    
    \hline
\end{tabular}}
\caption{\label{neutronic-cycles} Standard fuel cycles. They typically
produce neutrons at high rates but are better suited for  
thermal nuclear fusion concepts.}
\end{center}
\end{table}
Advanced fuel cycles are summarized in table \ref{aneutronic-cycles}.
They have larger Gamov energies $\epsilon^{kl}_G$ than the neutronic
fuel cycles. Hence, the cross sections are very small at low
energies. At high energies, however, aneutronic fuel cycles become
attractive as well. In particular, boron is a material that can be
nano-fabricated and doped to form an integrated
nano-accelerator. Protons, deuterons, and tritium can be
implanted into boron.
\begin{table}[ht]
\begin{center}
{\small
   \begin{tabular}{|l|c|c|c|}
    \hline
    advanced fuels & $\epsilon^{kl}_f$ MeV & $S_{kl}$ keV barn & $\sqrt{\epsilon^{kl}_{G}} \; \sqrt{\text{keV}}$
    \\
    \hline
    $\ce{D}+\ce{^3He} \rightarrow \ce{^4He} +\ce{p}$ & $18.35$ & $5.9 \cdot 10^3$ & $68.75$\\
    $\ce{p}+\ce{^6Li} \rightarrow \ce{^4He} +\ce{^3He}$ & $4.02$ &
                                                                   $5.5
                                                                   \cdot
                                                                   10^3$ &$87.2$ \\
    $\ce{p}+\ce{^7Li} \rightarrow 2\ce{^4He}$ & $17.35$ & $80.0$
                            &$88.11$ \\
    $\ce{p}+\ce{^{11}B} \rightarrow 3\ce{^4He}$ & $8.68$ &
                                                                   $2.0
                                                                   \cdot 
                                                                   10^{5}$
                            &$150.3$ \\
    \hline
   \end{tabular}}
\caption{\label{aneutronic-cycles} Advanced fuel cycles. They
typically produce less neutrons but have larger Gamov energies.
Hence, cross sections are small at low energies making the fuel
cycles difficult to trigger in a thermal context.}
\end{center}
\end{table}

\begin{figure}[ht]
\begin{center}
\includegraphics[width=70mm]{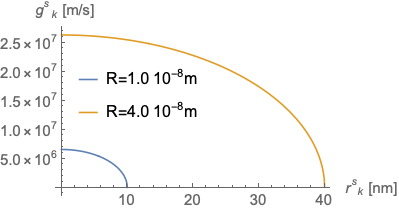}
\end{center}
\caption{\label{exit_velocity1} Exit velocities according to 
(\ref{exit_velocity}) as a function of initial radii. The nano-rod 
density is $n_l \approx 5 \cdot 10^{29} \, \text{m}^{-3}$.}
\end{figure}

We obtain
\begin{eqnarray}
\label{approximate-reactivity}
&&\hspace{-0.5cm}\sum_s \alpha^s_k \,  \int^{\infty}_{t^s_k} dt \, 
g^{s}_k(t) \, \sigma^{kl}_R \left( g^{\, s}_k(t) \right) \approx
\sigma^{kl}_0 \, \sum_s \alpha^s_k \, \frac{g^{s}_k\left(
t^s_k\right)}{\nu^s_{ke}} \, .
\end{eqnarray}
The exit velocities as a function of initial radius $r^s_k$ are shown
in Fig. \ref{exit_velocity1}.
For appropriate nano-structures the exit velocity spread of 
the light ions $k$ can be very small. With the help of an average
exit velocity $g^{av}_k$  we obtain 
\begin{eqnarray}
\label{conversion-fraction2}
\eta^{kl}&=&\frac{n_l \,  {\cal R}_k \,
\sigma^{kl}_0}{1+ n_l \, {\cal R}_k \, \sigma^{kl}_0} \, , \quad 
{\cal R}_k \approx \frac{g^{av}_k}{\nu_{ke}} \, .
\end{eqnarray}
The parameter ${\cal R}_k$ is the range of the fuel constituent $k$. Since the
resistivities scale like the density $n_l$ and the conversion fraction
contains the product $n_l \, {\cal R}_k$, there is no dependence on
the density $n_l$ within the model outlined here. According to
(\ref{collision-frequencies}) and (\ref{conversion-fraction2}) the
conversion efficiency $\eta^{kl}$ limited by
\begin{eqnarray}
\label{eta-limit}
&&\eta^{kl} <  \frac{\frac{4 \pi \epsilon^2_0 m^2_{ke} c^3 \, g^{av}_k \, 
\sigma^{kl}_R}{\alpha \, e^2_k e^2_e \, \ln \Lambda_{ke}} }{1+ \frac{4 
\pi \epsilon^2_0 m^2_{ke} c^3 \, g^{av}_k \, \sigma^{kl}_R}{\alpha \, 
   e^2_k e^2_e \, \ln \Lambda_{ke}} } \, ,
\end{eqnarray}
where $0 < \alpha < 1$ is the fraction of electrons divided by the
number of ions in the volume of consideration. For the parameters
\begin{eqnarray}
&&\alpha \approx 1 \, , \quad \ln \Lambda_{ke} \approx 5 \, , \quad
\sigma^{kl}_R \approx 10^{-28} \, \text{m}^2 \, , \nonumber
\end{eqnarray}
we have $\eta^{kl} < 0.02$. In the context there are many experiments
dealing with pitcher - catcher configurations falling short of the
approximate limit (\ref{eta-limit}). The convertor proposed here
represents an integrated pitcher - catcher configuration. It promises
to be an order of magnitude more efficient than traditional pitcher -
catcher configurations. 

For simple situations $\eta^{kl}$ depends on the $n_l \, {\cal R}_k$
product while $n_l$ and ${\cal R}_k$ are not independent of each other
as (\ref{eta-limit}) implies. To obtain a feeling for densities and
ranges required for large conversion fractions we consider
$\ce{p^{11}B}$ and $\ce{DT}$.

\begin{figure}[ht]
\begin{center}
\includegraphics[width=70mm]{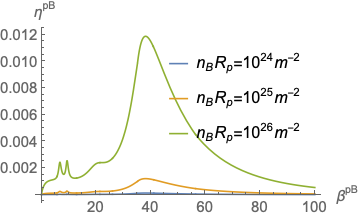}
\end{center}
\caption{\label{conversion-fraction_pB} The expected conversion
fraction $\eta^{pB}$ as a function of $\beta^{pB}_{av}$ and $n_B \, {\cal
R}_p$ based on the cross section given in
Fig. \ref{pB_cross_section_paper}. Please note that according to
(\ref{conversion-fraction2}) density and range are inversely proportional
to each other. Values for the product  $n_B \, {\cal R}_p$
can be obtained from (\ref{eta-limit}).}
\end{figure}

For an average fuel density of about $n_B = n_p
\approx 5.0 \cdot 10^{28} \, \text{m}^{-3}$, an average velocity of
$g^{av}_p \approx2.0 \cdot 10^7 \, \text{ms}^{-1}$ according to
Fig. \ref{exit_velocity1}, and a resistivity of $\nu_{pe}
\approx 10^{10} \, \text{s}^{-1}$ the range is ${\cal R}_p \approx 2 \cdot
10^{-3} \, \text{m}$ yielding $n_B \, {\cal R}_p \approx 10^{26} \, \text{m}^{-2}$.  
The conversion fraction $\eta^{pB}$ as a function
of $\beta^{pB}_{av}$ and $n_B \, {\cal R}_p$ is shown in
Fig. \ref{conversion-fraction_pB}. 

\begin{figure}[ht]
\begin{center}
\includegraphics[width=70mm]{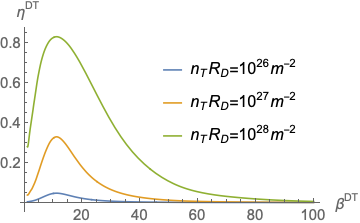}
\end{center}
\caption{\label{conversion-fraction_DT} The expected conversion fraction $\eta^{DT}$
  as a function of $\beta^{DT}_{av}$ and $n_T \, {\cal R}_D$. Please
  note that according to (\ref{conversion-fraction2}) density and range
  are inversely proportional to each other. Values for the
  product  $n_T \, {\cal R}_D$ can be obtained from (\ref{eta-limit}).}
\end{figure}

For deuteron and tritium ions embedded within a boron nano-rod
nano-acceleration is capable of generating energetic high density
deuterons and tritium ions, which can collide with each other.
The average density of the deuterium and tritium leaving the boron
nano-rods can be as high as $n_D \approx 5 \cdot 10^{28} \, \text{m}^{-3}$.
The exit velocity of $\ce{D}$ is assumed to be $10^7 \,
\text{ms}^{-1}$, while the resistivity is $\nu_{De} \approx 10^{10} \,
\text{s}^{-1}$. We obtain $n_T \, {\cal R}_D \approx 10^{26} \,
\text{m}^{-2}$. The conversion fraction $\eta^{DT}$ as a function of
$\beta^{DT}_{av}$ and $n_T \, {\cal R}_D$ is shown in
Fig. \ref{conversion-fraction_DT}.

Since laser technology has made vast progress in recent years the 
exploration of fusion enhancing micro-configurations operating close 
to solid fuel density and at extremely high fuel energies with or 
without auto-catalysm is of great interest. 

In section \ref{nonlinear_optics} some aspects of the nonlinear
electron optical properties of the micro-reactor are
addressed.

\section{Nonlinear optics} \label{nonlinear_optics}
Efficient embedded nano-acceleration of ions depends on specific
optical properties of the laser driver interacting with the nano-structures.

\begin{figure}[ht]
\begin{center}
\includegraphics[width=80mm]{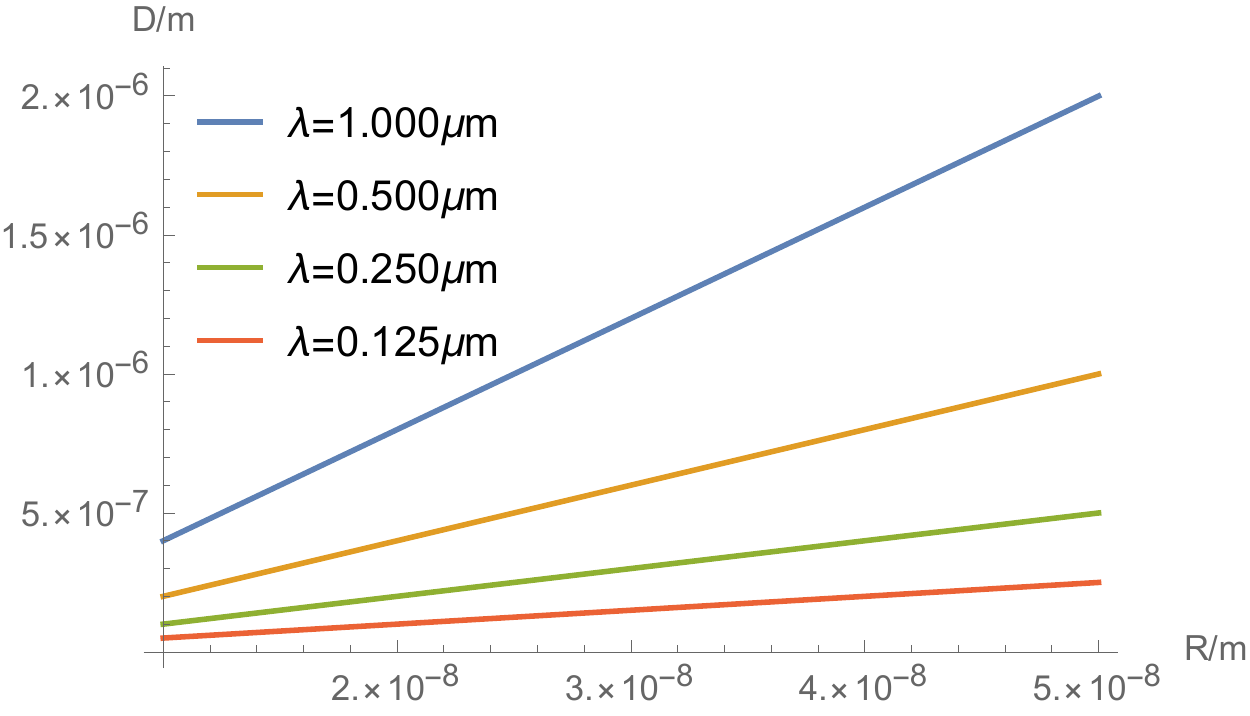}
\end{center}
\caption{\label{D-vs-rod-radius} Inter-rod gap $D$ versus nano-rod
radius $R$ for $\ce{p^{11}B}$. The fuel is assumed to be half ionized.}
\end{figure}

The lower threshold for the electric field strength required to ionize
the nano-rods to the charge density $e_l n_i$ is approximately
\begin{eqnarray}
&&E \ge \frac{R}{e_e} \, \left( m_e \, \omega^2 + e_e \, C_l \right) \, ,
\end{eqnarray}
where $\omega$ is the laser frequency. The implication for the laser intensity is 
\begin{eqnarray}
I_c &=& \frac{1}{2} \epsilon_0 \, c \, E^2 
   \ge  \frac{\epsilon_0 c \, R^2 \, \left( \frac{4 
   \pi^2 c^2 \, m_e}{\lambda^2} + e_e \, C_l \right)^2}{2 \,e^2_e} \, ,
\end{eqnarray}
where $\lambda$ is the laser wavelength. The gap $D$ between the
nano-rods is estimated from the critical plasma density for a given
$\lambda$ of the laser as
\begin{eqnarray}
D \ge \sqrt{\frac{e^2_e \, n_i \, R^2}{4 \pi \epsilon_0 \,
m_e c^2}} \, \lambda \, . 
\end{eqnarray}
Figure \ref{rod-radius-vs-intensity} below shows the approximate
radius $R$ of the rods for various wavelengths $\lambda$ for half
ionized $\ce{p^{11}B}$ required for the relative energy of $\epsilon^{pB}_g
\approx 0.5 \, \text{MeV}$ between protons and boron ions. 
\begin{figure}[ht]
\begin{center}
\includegraphics[width=80mm]{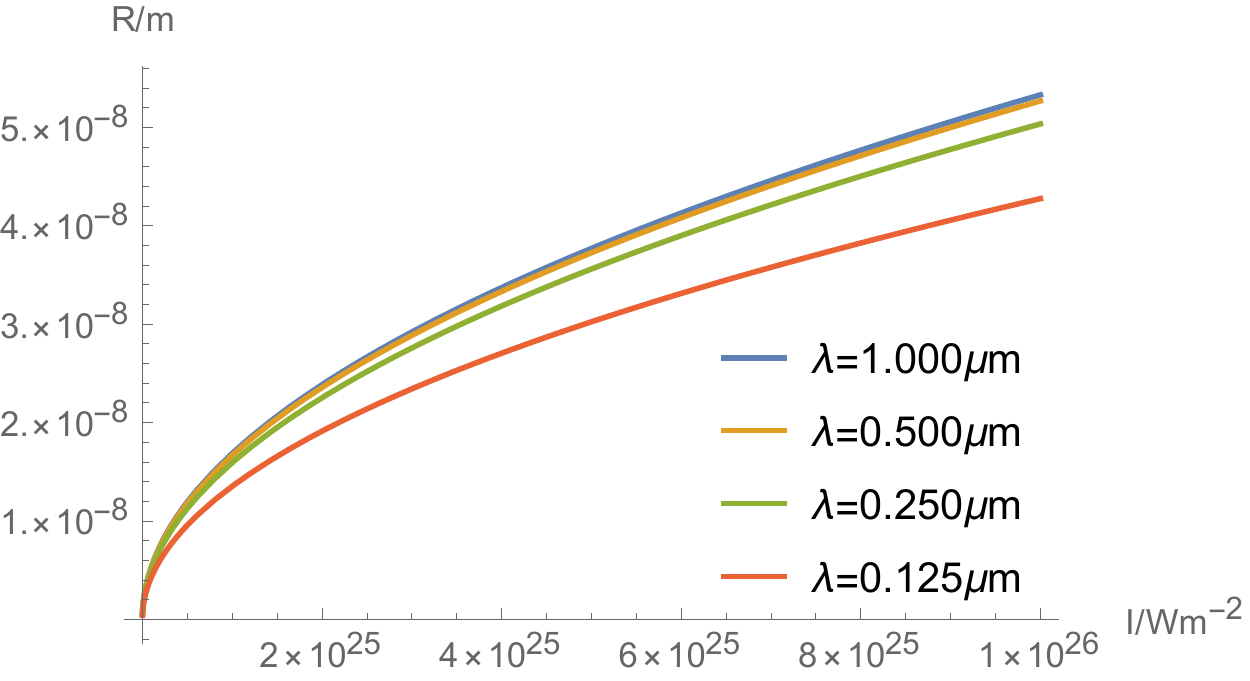}
\end{center}
\caption{\label{rod-radius-vs-intensity} Nano-rod radius $R$ versus
laser intensity for a relative energy of $\epsilon^{pB}_g \approx 0.5 \,
\text{MeV}$ between protons and boron ions for various laser
wavelengths for $\ce{p^{11}B}$. The fuel is assumed to be half ionized.}
\end{figure}
For a given rod radius $R$ and half ionized $\ce{p^{11}B}$ the
required gap $D$ between nano-rods for stable laser pulse propagation
is shown in Fig. \ref{D-vs-rod-radius}.

The gap $D$, the radius $R$, and the charge state $n_i$ of the rods
can be engineered to match the laser driver for optimal
nano-acceleration of ions. Within the scope of the model it is the
goal to convert a large share of the external laser energy into ionic
motion. Laser-optical instabilities have to be avoided.

Since the required intensities and charge densities are high, the power
of the required laser pulses might exceed the critical power for
self-focusing. However, self-focusing is suppressed for
sufficiently short laser pulses with $L\le \lambda_p$ according to
\cite{esarey1997self,sprangle1992propagation,faure2002effects}, where
$\lambda_p=2\pi/k_p$ is the plasma length. We have
\begin{eqnarray}
 &&P_{c,sp} \approx \frac{2 \, P_c}{k^2_p \zeta^2} \gg P_c \, , \\
&&P_c \approx 17 \, \frac{\omega^2}{\omega^2_{pe}}  \text{GW} \, ,
\end{eqnarray}
where $\zeta$ is the pulse length in the laser pulse frame. The time
required to ionize the nano-rods increases $P_{c,sp}$ further and
hence allows for laser pulses with $L > \lambda_p$.

The impact of secular electropmagnetic fields on $\eta^{kl}$ is  
neglected in the present paper. Since ultra-short laser pulses are
capable of capturing ionizing electrons from the nano-rods, large
electronic currents and return currents along the $z$-axis are
generated which produce secular electric and magnetic fields.

Neglecting collisional and radiative resistivities in
(\ref{transport-eqn-fusion2}), we have for the electronic fluid
\begin{eqnarray}
\label{mhd1}
&&\hspace{-1cm} \frac{\partial n_e}{\partial t} + \frac{\partial}{\partial \vec x} \cdot \left( n_e \, \vec{v}_e 
   \right) =0 \, , \\
\label{mhd2}  
&&\hspace{-1cm}\left( \frac{\partial}{\partial t}  + \vec{v}_e \cdot
   \frac{\partial}{\partial \vec x} \right) \, \vec{v}_e 
= -\frac{q_e}{m_e} \, \left( \vec{E} + \vec{v}_e \times \vec{B} \right)  + \frac{q_e}{m_e n_e} \,
   \frac{\partial P_e}{\partial \vec x}  \, ,
\end{eqnarray}
where
\begin{eqnarray}
\label{mhd3}
&&\frac{\partial \vec E}{\partial t} =
   \frac{\partial}{\partial \vec x} \times \vec B -
   \frac{1}{\epsilon_0 c^2} \vec j_e \, , \\
\label{mhd4}
&&\frac{\partial \vec B}{\partial t} = -
   \frac{\partial}{\partial \vec
   x} \times \vec E \, , \\
\label{mhd5}
&&\frac{\partial}{\partial \vec x} \cdot \vec E = \frac{1}{\epsilon_0}
\, \rho \, , \\
\label{mhd6}
&&\frac{\partial}{\partial \vec x} \cdot \vec B = 0 \, .
\end{eqnarray}
The parameter $P_e$ is the electronic plasma pressure. The electric and
magnetic fields in (\ref{mhd1}) - (\ref{mhd6}) are slowly varying
after the laser pulse has left, as simulations confirm. Assuming $d\vec
v_e/dt \approx 0$ and piecewise constant $n_e$ and $j_e$ we obtain
with the help of (\ref{mhd2})
\begin{eqnarray}
  &&n_e \, \vec E + n_e \, \vec{v}_e \times \vec{B}  - \frac{\partial P_e}{\partial \vec x}
     \approx 0 \, , \\
  && \frac{\partial \vec B}{\partial t}  =
   \frac{\partial}{\partial \vec x} \times \left( \vec v_e \times \vec B \right)
     \approx 0 \, , \\
  &&\frac{\partial}{\partial \vec x} \times \vec E \approx 0 \, , \\
  && \frac{\partial}{\partial \vec x} \times \vec B \approx
     \frac{1}{\epsilon_0 c^2} \vec j_e \, .
\end{eqnarray}
The strength of the magnetic fields can be estimated if the electronic
current densities are known. We approximate
\begin{eqnarray}
\label{stabilizing_B}
 &&\vec B \left( \vec r, t \right) \approx
     \left\{
     \begin{array}{ll}
       \frac{j_e r}{2\epsilon_0 c^2} \, \vec e_{\phi} \, , & r \le R_L \\
       0 \, , & r > R_L \\
     \end{array}
  \right. \, , 
\end{eqnarray}
where $j_e=e_en_ev_e$ is the strength of the electronic current density and
$R_L$ is the effective laser pulse radius. The electric field
associated with $j_e$ and $B$ is according to (\ref{mhd2})
\begin{eqnarray}
\label{stabilizing_E}
\vec E\left( \vec r, t \right) &\approx& - \frac{1}{e_e n_e}
\vec j_e \times \vec B(r) + \frac{1}{m_e} \,
   \frac{\partial P_e}{\partial \vec x} \, .
\end{eqnarray}
The discussion of the impact of secular electromagnetic fields on
$\eta^{kl}$ is beyond the scope of the present paper.

In section \ref{collisions_radiation} we discuss radiative energy
loss to obtain a feeling for the time scales involved.

\section{Radiative energy loss} \label{collisions_radiation}
Since the transport framework (\ref{kl-system}) -
(\ref{current-densities}) is based on a BBGKY-hierarchy up to binary
correlation order, we have radiative contributions to the equations of
motion of charged particles by mean-field radiation which is
traditionally called radiation reaction \cite{bild2019radiation} and
by radiative collisions in binary correlation order. The principal
calculation of the latter is conceptually outlined in
\cite{king2013trident}. They are obtained with the help of the
${\cal T}$-matrix in (\ref{eq:t-matrix}).

Both contributions modify the equations of motion of an
electron. While electrons are subject to self-radiation, binary level
radiative collisions, and the impact of external fields, ions are
mainly subject to radiation-free collisions and the impact of
secondary collective fields, as outlined in (\ref{solution_inv}) and
(\ref{rate2}) - (\ref{rate3}). Their reactive dynamics is given by
(\ref{rate1}) - (\ref{rate3}) or, on a very detailed level, by the
underlying kinetic equations (\ref{kl-system}) -
(\ref{current-densities}) that contain radiation reaction
\cite{bild2019radiation}. The description of radiation reaction based on
the approach in \cite{bild2019radiation} to a quantum level is in
preparation.

Ions mainly lose their energy via collisions with electrons. As long
as electrons are hot they cannot collide efficiently with cold
ions. The interaction of the electrons with the external laser driver
accelerates them to the speed of light. However, due to radiative energy loss
electrons lose their initial energy and eventually collide with
ions, thus draining the energy contained in the ionic subsystem.

The radiation loss per single electron can be estimated to be
\cite{bild2019radiation,nikishov1964quantum}
\begin{eqnarray}
&&\frac{dp^{\mu}}{d\tau} \approx \frac{2 \tau_0}{3} \, \frac{m_e}{c^2}
   \, a^{\nu} a_{\nu} \, u^{\mu} \, , \quad \tau_0 =
   \frac{e^2}{4\pi \epsilon_0 \, m_e \, c^3} \, .
\end{eqnarray}
The radiation power loss per electron is given by
\begin{eqnarray}
&&\frac{d \left( c \, p^0 - m_e c^2 \right)}{dt} \approx \frac{2 \tau_0}{3} \, \frac{m_e}{c^2}
   \, a^{\nu} a_{\nu} \, c^2 \, ,
\end{eqnarray}
where
\begin{eqnarray}
&&\frac{dx^{\nu}}{d\tau} = u^{\nu} \, , \quad \frac{du^{\nu}}{d\tau} = a^{\nu}
   \approx \frac{e}{m_e} \, F^{\nu \alpha} \, u_{\alpha} \, .
\end{eqnarray}
According to Landau and Lifshitz \cite{ll4_german} we have
\begin{eqnarray}
a^{\nu} \, a_{\nu} &\approx& -\frac{m^2_e c^6}{\hbar^2} \, \chi^2_e \, ,
\quad \chi_e = \frac{e\hbar}{m^3_ec^3} \, \sqrt{-\left( F^{\nu \alpha} \,
p_{\alpha}\right)^2} \, .
\end{eqnarray}
The strongest field in the fast micro-reactor is the laser field. We assume
\begin{eqnarray}
\chi_e \approx \frac{\gamma E}{E_s} \approx 10^{-5} \, ,  \quad E_s =
  \frac{m^2_e c^3}{e \hbar} \approx 10^{18} \,
  \frac{\text{V}}{\text{m}} \, .
\end{eqnarray}
This implies for the electronic radiation loss power density
\begin{eqnarray}
&&I_{cl} \approx \frac{1}{6 \pi} \, \frac{e^2 m^2_e c^3 n_e}{\epsilon_0 \hbar^2} \, \chi^2_e
\approx 3.2 \cdot 10^{-7} \, n_e \, \chi^2_e \,
   \frac{\text{J}}{\text{ps}} \, .
\end{eqnarray}
It is also possible to calculate the quantum corrections to radiation
reaction including the impact of the nonlinear Compton effect according to the
papers by Ritus and Nikishov \cite{nikishov1964quantum}. The emitted
integrated radiation loss power density for $\chi_e \ll 1$ according to
\cite{nikishov1964quantum} is
\begin{eqnarray}
&&I_{RN} \approx I_{cl} \, \left( 1 - \frac{55 \sqrt{3}}{16} \, \chi_e
   + 48 \, \chi^2_e \pm ... \right) \le I_{cl} \, .
\end{eqnarray}
Binary correlation order radiative processes cannot drain electronic
energy faster than the total energy of a radiation electron multiplied
by the binary collision frequency 
\begin{eqnarray}
&&\nu_{ke} \approx \frac{e^2_k e^2_l \, n_l}{4 \pi \epsilon_0 m^2_{ke} \,
   \text{v}^3_{ke}} \, \ln \Lambda_{ke} \, , \quad \text{v}_{ke} = \left|
   \vec v_k - \vec v_e \right| \, , 
\end{eqnarray}
where $k$ can also represent an electron. We estimate
approximately $\nu_{ke} \approx \left( 10^{9} - 10^{11} \right) \,
\text{s}^{-1}$ and hence electronic energy loss in the fast micro-reactor
at near solid density via collisions takes about $10 - 1000 \, \text{ps}$.

\section{Summary} \label{summary}
We propose an emergent micro-reactor concept based on very small
nano-structures interacting with powerful short laser pulses for
triggering nuclear fusion reactions appropriate for a range of nuclear
fuels. It is driven by advanced ultra-short ultra-high energy laser
pulses in the UV to the VUV wavelength range. The embedded
nano-structures are quite small and represent an integrated
nano-accelerator.

The micro-reactor is a concept promising efficient nuclear fuel
conversion at near solid fuel density, which can be tested
experimentally. The reactor proposed here should be far more
efficient than traditional pitcher - catcher configurations. 

With the help of the analysis of conversion fractions $\eta^{kl}$
fusion enhancing configurations in emergent micro-reactors can
be addressed. In the paper the conversion fraction $\eta^{kl}$ for an
integrated pitcher - catcher configuration has been analyzed.

Due to the availability of modern lasers the scope of
fusion enhancing configurations is much wider than the one of
classical confinement configurations discussed in the context of
traditional fusion devices. 

The present paper represents an introduction into a potentially
powerful micro-reactor concept. We will follow up with this paper by a
number of more detailed papers that focus on advanced fusion enhancing
configurations and the associated limiting conversion fractions.

\section{Acknowledgements}
The present work has been motivated by the Marvel Fusion
GmbH. 




  \bibliographystyle{elsarticle-num} 
  \bibliography{literatur_eqn_motion}





\end{document}